\documentclass[preprint]{aastex}
\shortauthors{Wyder}
\shorttitle{Star Formation History of NGC 6822}
\begin{document}
\title{The Star Formation History of NGC 6822}
\author{Ted K. Wyder}
\affil{California Institute of Technology, Mail Code 405-47, 
1200 E California Blvd, Pasadena, CA 91125}
\email{wyder@srl.caltech.edu}

\begin{abstract}

Images of five fields in the Local Group dwarf irregular galaxy NGC
6822 obtained with the {\it Hubble Space Telescope} in the F555W and
F814W filters are presented. Three of the fields lie mostly within the
central bar of NGC 6822 where most of the sites of current star
formation are concentrated while two other fields sample more of the
outer regions of the galaxy.  Photometry for the stars in these images
was extracted using the Point-Spread-Function fitting program
HSTPHOT/MULTIPHOT.  The resulting color-magnitude diagrams reach down
to $V\approx26$, a level well below the red clump, and were used to solve
quantitatively for the star formation history of NGC 6822.  Assuming
that stars began forming in this galaxy from low-metallicity gas and
that there is little variation in the metallicity at each age, the
distribution of stars along the red giant branch is best fit with star
formation beginning in NGC 6822 12-15 Gyr ago. The best-fitting star
formation histories for the old and intermediate age stars are similar
among the five fields and show a constant or somewhat increasing star
formation rate from 15 Gyr ago to the present except for a possible
dip in the star formation rate from 3 to 5 Gyr ago.  The main
differences among the five fields are in the higher overall star
formation rate per area in the bar fields as well as in the ratio of
the recent star formation rate to the average past rate. These
variations in the recent star formation rate imply that stars formed
within the past 0.6 Gyr are not spatially very well mixed throughout
the galaxy. The star formation histories in conjunction with a galaxy
evolution code are used to infer the evolution of the integrated
absolute magnitude of NGC 6822 as a function of look-back time and
redshift. The results indicate that galaxies with star formation
histories similar to that determined for NGC 6822 would have an
absolute magnitude of $M_B=-15.7$ at a redshift $z=0.6$, well below
the detection limit of the current redshift surveys, and therefore are
most likely not contributing to the evolving galaxy population
detected in deep redshift surveys.

\end{abstract}


\section{Introduction}

The dwarf galaxies of the Local Group provide excellent laboratories
with which to study galaxy evolution. Dwarf galaxies are in general
useful in studying stellar populations and galaxy evolution in the
absence of the complicating effects of density waves common in giant
disk galaxies.  Dwarf galaxies have been found at a variety of
distances from giant galaxies, allowing us to investigate the
importance of environment in their evolution. Their low metallicities
allow us to study how galaxies form stars under conditions similar to
that at high redshift in larger galaxies. The proximity of Local Group
galaxies means that we can study them in greater detail that for any
other galaxies. For recent comprehensive reviews of the properties of
Local Group galaxies, see \citet{hod89}, \citet{mat98}, and
\citet{vdb00}, among others.

Detailed studies of nearby dwarf galaxies are also relevant to
understanding the nature of the faint blue galaxies detected in deep
imaging and redshift surveys \citep{ell97}.  The redshift surveys
described by \citet{lil95} and \citet{ell96} show that the faint blue
galaxies lie at redshifts of $0.2<z<1.0$,
have strong [OII] emission indicative of active star formation, and
contribute to a steepening of the faint end slope of the luminosity
function. One model that has been proposed to explain these
observations is that the evolving population in these surveys is
comprised of dwarf galaxies that form all of their stars in a single
burst of star formation lasting $\sim 10^7$ years at a redshift of $z
\sim 1$ \citep{bab92,bab96}. While none of the star formation
histories (SFHs) of Local Group dwarf galaxies yet observed fit exactly this
scenario, it may be possible that some of the Local Group dwarfs 
may be the low redshift counterparts to
the evolving population detected in the redshift surveys. By
determining the SFHs of as many nearby galaxies
as possible and comparing them with the properties of distant
galaxies, we should be able to come to a better understanding of the
nature of the faint blue galaxies and their low redshift counterparts.

With a few exceptions, the dwarf galaxies of the Local Group can be
classified into early type dwarf ellipticals and dwarf spheroidals and
late type dwarf irregulars. The dwarf ellipticals and dwarf
spheroidals are gas poor, have no current star formation and tend to
be concentrated close to M31 and the Milky Way while the dwarf
irregulars are gas rich, are currently forming stars and have a more
uniform distribution throughout the Local Group \citep{mat98,vdb00}.
\citet{mat98} presented a summary of observations of a number of Local
Group galaxies which illustrated the variety of SFHs observed. While
none of the dwarf galaxies share the exact same SFH, there are a few
trends that have emerged.  While it had once been assumed that all
dwarf spheroidals contained exclusively old populations, it has become
clear that there are many dwarf spheroidals, such as Carina
\citep{sme94}, that have undergone multiple star formation episodes
resulting in a dominant population of intermediate age stars. On the
other hand, there is increasing evidence that many dwarf irregular
galaxies, such as NGC 6822 \citep{gal96a} and IC 1613 \citep{col99},
contain significant intermediate age and old stellar
populations. The only dwarf irregular known to be composed
of almost entirely young ($< 1$ Gyr old) stars is Leo A
\citep{tol98} while Draco and Ursa Minor are the only dwarf
spheroidals known which contain only very old stars \citep{gri98,mig99}.
Despite the substantial progress made in understanding
the evolution of dwarf galaxies, the nature of the relationship
between dwarf spheroidal and dwarf irregular galaxies remains unknown.

NGC 6822 in particular is a Local Group dwarf irregular galaxy that
has been the subject of numerous observational studies at a variety of
wavelengths.  In his historic paper, \citet{hub25} presented his
discovery of Cepheids in this galaxy which made NGC 6822 the first
object recognized as a truly extragalactic object well outside the
Milky Way. Since then, the distance to NGC 6822 has become fairly well
established in that it has been measured both using the Cepheid
period-luminosity relation and the tip of the red giant
branch. \citet{mca83} obtained a distance modulus of
$(m-M)_0=23.47\pm0.11$ based upon their $H$-band photometry of nine
Cepheids in NGC 6822. Using $BVRI$ photometry to recalibrate the
Cepheid light curves originally measured by \citet{kay67} and assuming
a distance modulus to the LMC of 18.5, \citet{gal96a} derived a
distance modulus of $23.49\pm0.08$ (500 kpc), the value that I have
adopted in the analysis presented in this paper.  These Cepheid
distances are consistent with the tip of the red giant branch distance
moduli of 23.46 and 23.40 obtained by \citet{lee93} and
\citet{gal96a}, respectively.

Since it lies at a rather low Galactic latitude of $b=18.4^{\circ}$,
the foreground reddening toward NGC 6822 is significant and has a
value of $E(B-V)=0.24$ \citep{sch98}.  The optical structure of NGC
6822 consists of a bar oriented roughly north-south containing most of
the sites of current star formation while the outer regions
surrounding the bar contain a population of fainter, redder and most
likely somewhat older stars \citep{hod77,hod91}.   Studies of OB stars
in NGC 6822 indicate that the reddening increases from the foreground
value in the outer regions to values of $E(B-V)=0.45$ in the vicinity
of the bar \citep{mas95}.  There are numerous \ion{H}{2} regions
concentrated in the bar which have been studied most recently by
\citet{hod88,hod89} and \citet{col95}.  Spectra of some of the
\ion{H}{2} regions in NGC 6822 indicate oxygen abundances
substantially less than solar
\citep{pei70,smi75,leq79,pag80,ski89b}. In particular, \citet{ski89b}
derived an oxygen abundance of $O/H=(1.6\pm0.4)\times10^{-4}$ which
corresponds to an overall metallicity of 20\% solar or a fraction by
mass of metals of $Z=0.004$.  This places NGC 6822 near the average
metallicity-luminosity relationship for dwarf galaxies \citep{ski89a}.
NGC 6822 contains a large reservoir of \ion{H}{1} gas that extends
over a region much larger than the optical diameter of $D_{25}=15.5\arcmin$
\citep{dav72,rob72,got77,deb00}. Using their best-fitting inclination
of $69^{\circ}\pm3$, \citet{got77} derived a total mass for NGC 6822
from their \ion{H}{1} rotation curve of $1.4\times10^9$ M$_{\odot}$
within a radius of $18\arcmin$, or 2.6 kpc. The far infra-red emission
in NGC 6822 as detected by the IRAS satellite has been analyzed by
\citet{gal91} and \citet{isr96}. Despite its low metallicity, CO
emission associated with molecular clouds in NGC 6822 has been
detected \citep{wil92a,wil94,isr97}.

Beginning with \citet{kay67}, there have been numerous studies of the
stellar content of NGC 6822. \citet{arm85} discovered 12 Wolf-Rayet
stars in NGC 6822 while \citet{hoe86} and \citet{wil92b} have
determined the blue star luminosity function. The OB associations in
this galaxy have been cataloged and studied by \citet{hod77} and
\citet{wil92b}. Many carbon stars were discovered by \citet{coo86}
in two fields at either end of the bar, providing the first evidence
for a significant intermediate-age population in NGC 6822.
One of the star clusters in NGC 6822, Hubble VII, is
likely similar in age and metallicity to the Milky Way globular
clusters while the clusters Hubble VI and Hubble VIII are probably
similar to the ``populous clusters'' observed  in the Magellanic
Clouds \citep{coh98,wyd00,cha00}. In addition there are numerous smaller
open clusters cataloged by \citet{hod77}.

Recently, there have been two attempts to quantify the SFH of NGC 6822
based upon color-magnitude diagrams (CMDs) derived from ground-based
data. Using $BV$ photometry of two fields in the vicinity of the bar
reaching down to $V\simeq23.5$,  \citet{mar95} derived information
about the recent ($<1$ Gyr) SFH by comparing the observed CMD with
synthetic diagrams. They found evidence for a more continuous star
formation rate (SFR) in one of the fields while the other is better fit
with a SFH containing two star formation episodes each lasting a few
hundred Myr.  Gallart et~al. (1996a,b,c) investigated the SFH using
$VI$ CMDs reaching down to $I\simeq22$ that covered an $11.2\arcmin
\times 10.4\arcmin$ area centered on the bar. By comparing the
observed CMDs with model diagrams, \citet{gal96a} found that NGC 6822
most likely began forming stars 12-15 Gyr ago from low metallicity
gas. The observed CMDs could also be consistent with star formation
beginning only 6-9 Gyr ago if prompt metal enrichment occurred in the
galaxy. Over the past few Gyrs, a SFR that is constant or declining
fits the observations. For the most recent SFH, \citet{gal96b} found
evidence for an enhancement in the field SFR in the last 100-200 Myr
with the enhancement being greater in the center of the bar than in
the outer regions and greater still at the ends of the
bar. \citet{hod80} also found evidence for an enhancement in the
cluster formation rate 75-100 Myr ago using cluster ages determined by
the magnitude of the brightest star in each cluster.

The main goal of this paper is the analysis of the stellar populations
and star formation history of NGC 6822 based mainly upon five 
{\it Hubble Space Telescope (HST)}
pointings within the galaxy. The higher resolution {\it HST} data
of the fields presented here is essential in coming to a better
understanding of the evolutionary history of this galaxy and are a
natural  compliment to the previous more spatially complete but
shallower ground-based data. With this new data, the variation in the
star formation history with position in NGC 6822 can be investigated
as well as the age of the oldest stars. Knowledge of the star
formation history of NGC 6822 is an essential step in comparing the
evolutionary history of NGC 6822 with other nearby dwarf galaxies as
well as understanding the nature of the evolving population of
galaxies detected in deep redshift surveys.  This paper is
organized as follows.  \S2 describes the data and photometry of the
five fields. In \S3, I describe the methods used to extract information
about the star formation history of each field while the results of
these analyses are presented in \S4. The main conclusions of this paper
are summarized in \S5.

\section{Observations and Photometry}

\subsection{Observations}

The observations presented here consist of images of five fields in
NGC 6822 taken with the WFPC2 aboard {\it HST}. The locations of these
fields in NGC 6822 are shown in Figure \ref{dss}. In four of the fields, the
PC chip is centered on one of the clusters H~IV, H~VI, H~VII or
H~VIII originally identified by \citet{hub25} while the fifth field is
centered on the open cluster C25 from the list in \citet{hod77}.  All
five fields were imaged through the F555W and F814W filters
which are similar to the standard $V$ and $I$ filters,
respectively. Two exposures per field in each filter
were obtained to assist in removal of cosmic rays. I obtained the
images from the Space Telescope Science Institute after the data were
processed by the standard WFPC2 calibration pipeline described in
detail in \citet{hol95a}. The Hubble cluster fields are a part of
{\it HST} program 6813 proposed by Paul Hodge (PI), Mario Mateo and
Toby Smith while the C25 observations are a part of {\it HST} program
8314 proposed by Paul Hodge (PI), Eugene Magnier, Andrew Dolphin and
Rupali Chandar. A summary of the observations is shown in Table \ref{obs_log}.
The analysis of the clusters appearing in the PC images has already been
presented in \citet{wyd00}.

\subsection{HSTPHOT/MULTIPHOT photometry}

To obtain photometry for the stars in these images, I relied on a new
PSF-fitting photometry package, HSTPHOT \citep{dol00b}, kindly provided
to me by the author. HSTPHOT is similar in many respects to other
PSF-fitting stellar photometry programs such as DAOPHOT/ALLSTAR
\citep{ste94} and DoPhot \citep{sch93}. HSTPHOT is designed specifically
for use with {\it HST} WFPC2 data and relies on a library of Tiny Tim
PSFs for the PSF-fitting. The PSFs are sampled on a $200\times200$
pixel grid on each chip to account for spatial variations in the
stellar shapes.  The main innovations of this software are that it
subsamples the PSF on a $5 \times 5$ grid within each PC pixel and on
a $10 \times 10$ grid within each WF pixel and accounts for the
subpixel variation in the quantum efficiency.  (See Dolphin 2000b for
a detailed description of HSTPHOT). An earlier version of this software
was used by \citet{hod99} and \citet{dol00a} to analyze WFPC2
observations of the Local Group dwarf galaxy WLM (DDO 221).

Before performing the photometry, I averaged together the two
individual exposures per cluster taken in each filter using a routine
included in the HSTPHOT package, {\it crclean}.  Cosmic rays are
rejected in this program using an algorithm nearly identical to that
implemented in the IRAF task {\it crrej}. Before performing the photometry,
the data quality files that accompany the data were used to replace the
value in bad pixels with $-100$ DN so that they will not be considered
when extracting the photometry.

The HSTPHOT package includes the program MULTIPHOT which actually
performs the PSF-fitting. MULTIPHOT differs from other photometry
programs in that it solves for the center and magnitudes for each star
in all of the input images in all filters simultaneously. I have used
MULTIPHOT to analyze the field stars in all five fields.  MULTIPHOT
modifies the PSF for each image to account for any focus variations
that may cause the real PSF to deviate from the Tiny Tim model.

The cosmic-ray cleaned images output by {\it crclean} in each filter
were input into MULTIPHOT. Stars were identified using a minimum
signal-to-noise threshold of 3.  In calculating a magnitude for each
star in each filter, MULTIPHOT corrects for the effects of the
geometric distortion on the photometry as well as the 34th row error.
The counts were also corrected for the charge-transfer-efficiency
(CTE) effect using the results presented in \citet{dol00c}. This CTE
correction used here is similar to that determined by \citet{whi99}.

Aperture corrections were determined automatically by MULTIPHOT by
measuring aperture magnitudes within a radius of $0.5\arcsec$ for
several of the brightest stars on each chip from an image in which all
of the other stars have been subtracted.  Then a mean aperture
correction for each chip and filter was determined by averaging the
differences between the PSF-fit and aperture magnitudes.  The standard
deviations of the aperture minus the PSF-fit magnitudes in all of the
fields are in the range of $0.02-0.05$ magnitudes. I conservatively
adopt this as the likely uncertainty in the aperture corrections and
it probably represents the dominant source of error in the overall
photometry zeropoint.

The instrumental F555W and F814W aperture-corrected magnitudes were
then calibrated and transformed to standard $V$ and $I$ using the
transformations presented in \citet{dol00b}. The color terms in that
paper are those originally determined by \citet{hol95b} while the
zeropoints have been redetermined via comparison of WFPC2 exposures of
the globular clusters $\omega$~Cen and NGC~2419 with ground-based
$UBVRI$ data. The $V$ and $I$ zeropoints assumed here differ from the
commonly used \citet{hol95b} zeropoints by $-0.009$ and 0.012
magnitudes in $V$ and $I$, respectively,  where the differences are in
the sense of \citet{hol95b} minus \citet{dol00b}. The new calibration
also incorporates new determinations of the gain for each chip which
result in zeropoint differences of $-0.044$, 0.007, $-0.007$ and
$-0.006$ mag relative to the \citet{hol95b} results for the PC, WF2,
WF3 and WF4, respectively.

MULTIPHOT calculates two parameters for each star, $\chi$ and {\it
sharp}, which are very similar to parameters output by DAOPHOT and can
be used to eliminate poorly fit stars as well as non-stellar objects.
$\chi$ is a measure of the overall goodness of the PSF fit while {\it
sharp} indicates whether the object may be a hot pixel or a background
galaxy.  An object that is completely flat would have a {\it sharp}
value of $-1$ while an object sharper than the PSF would have a
positive value.  In less crowded fields, well fit stars should have
$\chi<2.5$ \citep{dol00b}.  For the very crowded fields analyzed here,
I have adopted a somewhat larger $\chi$ limit to insure that the
photometry is as complete as possible. Any biases introduced by the
adopted $\chi$ cut should be accounted for by the artificial stars
tests described below.  Only stars with $\chi<6$ and $-0.5<${\it
sharp}$<0.5$ in both filters were retained for the final photometry
list. These same limits were used for all five fields. The resulting
$V, I$ color-magnitude diagrams for all stars detected in the WF chips
in each field are plotted in Figure \ref{obs_cmds}.  There are a total
of 34,285, 47,872, 45,312, 32,303 and and 27,823 stars detected in the
three WF chips in the H~IV, H~VI, H~VII, H~VIII and C25 fields,
respectively.  In each of the panels of Figure \ref{obs_cmds},  the
data are plotted as individual points in those parts of the diagrams
where the density of points is low.  However, in those areas of each
CMD where the density of stars is greater than 20 decimag$^{-2}$, the
data are plotted as contours. All of the CMDs throughout the remainder
of this paper are plotted in the same manner.

\subsection{Artificial Star Tests}

The analysis of the CMDs described in the remaining sections requires a
quantitative assessment of the observational effects on the observed
CMD. This has been done using the results of a series of artificial star
tests performed on each of the five fields with routines within the
MULTIPHOT program.  The artificial stars were given random magnitudes
and colors in the range $20<V<29$ and $-0.5<V-I<3.5$ for the four
Hubble cluster fields while the artificial stars in the C25 field were
chosen to have magnitudes in the range $18<V<28$ due to the shorter
exposure time. In each field the total number of stars added ranged
between $9\times10^5$ and $1.1\times10^6$ stars per field.
Positions for the stars were chosen to mimic the
distribution of light in each image.  The artificial stars were then
convolved with the appropriate PSF for its chip, filter and position
and the appropriate amount of noise added. Then the artificial stars
were detected and measured using the same algorithms as for the
observed data. Since the artificial stars follow the model PSF exactly
(with the addition of random Poisson noise), it is important to note
that these completeness tests only simulate the uncertainties
associated with the crowding, Poisson noise and sky determination and
would not be able to account for any uncertainties due to any possible
mismatch between the model and observed PSF.

An input artificial star was classified as recovered if it was
recovered in both $V$ and $I$ with $\chi<6$ and $-0.5<${\it
sharp}$<0.5$ in both images.  The results of the artificial star tests
for each field are shown in Figure \ref{fakestars}. Each panel of this
figure shows the fraction of artificial stars recovered as a function
of the input $V$ magnitude. The completeness results are plotted
separately for three $V-I$ bins. In each field, the completeness for
the red stars is greater than in the blue. At the bright end, the
completeness is about 95\% in all fields. For fainter magnitudes, the
completeness falls off more quickly with increasing magnitude for the
more crowded fields H~VI and H~VII than in the H~IV and H~VIII
pointings despite the nearly identical exposure times in all the
fields. 

\subsection{Comparison of H VI and H VII photometry}

As can be clearly seen in Figure \ref{dss}, the H~VI and H~VII fields
contain a significant amount of overlap, allowing me to check the
internal consistency of the photometry. The IRAF task {\it metric} was
used to convert from pixel position on a given chip to RA and Dec
coordinates on the sky for all of the stars measured in each
field. {\it Metric} relies upon a solution of the geometric distortion
in each field as well as the astrometry information in the image
headers. Blind application of the {\it metric} astrometry resulted in
few matches between the two fields. A comparison of the coordinates of
individual stars on both fields revealed that there were offsets in RA
and Dec between the {\it metric} astrometry for the two fields even
though the scatter was relatively small.  The offsets were also a
function of which pair of chips each star happened to fall upon and
ranged up to an offset of $0.37\arcsec$.  These offsets were used to
correct the {\it metric} astrometry. A final list of stars detected in
both fields was derived using a matching radius of $0.05\arcsec$.

If the artificial star tests described in the previous section provide
an adequate description of the photometric completeness for each
field, then the fraction of stars within the overlap area that are
detected in both the H~VI and H~VII fields should be equal to the
product of the completeness fractions in each field.  The fraction of
stars detected in the H~VII field that both lie within the H~VI
field-of-view and were detected in the H~VI field is plotted in Figure
\ref{h6vsh7_fraction} as the solid line. The fraction of stars
detected in both fields reaches 90\% at the bright end and begins to
fall rapidly for magnitudes $V>24$.  The dotted line in the figure is
the product of the completeness fractions for the H~VI and H~VII fields
derived from the results of the artificial star tests for stars with
$0.5<V-I<1.5$, the range of color where most of the stars in the CMD
are found. The agreement between these two curves indicates that the
completeness of the photometry in these two fields is entirely
consistent with the artificial star test results.

Plots of the magnitude differences, $\Delta V$ and $\Delta I$, divided
by the expected error in both $V$ and $I$, $\sigma_{V,expected}$ and
$\sigma_{I,expected}$, for stars in common between the two fields are
shown in panels (b) and (d) of Figure \ref{h6vsh7}.  The values of
$\sigma_{V,expected}$ and $\sigma_{I,expected}$ were defined to be the
quadrature sum of the error for each field, shown in panels (a) and
(c) of Figure \ref{h6vsh7} for $V$ and $I$, respectively.  These
photometric errors do not include any zeropoint, gain or aperture
correction uncertainties and are defined to be the standard deviations
of the difference between the input and recovered magnitudes for stars
in the artificial star tests. The solid lines in panels (b) and (d) of
the figure correspond to zero difference while the dotted lines mark
the $\pm 1 \sigma$ deviations from zero. While there are stars with
magnitude differences between the two fields as large as $4-5 \sigma$,
83\% and 85\% of the stars lie within $\pm 1$ standard deviation of
zero in the $V$ and $I$ data, respectively. The results plotted in
Figure \ref{h6vsh7_fraction} and \ref{h6vsh7} argue that the
artificial star tests provide a reasonable representation of the
internal photometric uncertainties and incompleteness in the data.

For the stars measured on both the H~VI and H~VII fields, an average
difference, $\Delta V$ and $\Delta I$, and scatter, $\sigma_V$ and
$\sigma_I$, for each chip combination was determined by taking the
mean difference  and standard deviation for stars with $20<V<23$ and
$19<I<22$. These average differences and standard deviations are
listed in Table \ref{h6vsh7tab}.  With one exception, the average
differences in magnitude agree to within 0.05 mag.  With the exception
of the WF vs. PC differences, the standard deviations lie in the range
$0.07-0.09$ and $0.06-0.07$ in $V$ and $I$, respectively.  These
zeropoint differences are similar to the spread mentioned earlier in
the aperture corrections for each field. Assuming that the relative
chip zeropoints and gain ratios as well as the CTE correction were
correctly determined by Dolphin (2000b,c), the most likely explanation
for the offsets seen between some chips and not others is the
uncertainty in the aperture corrections. Furthermore, there are fewer
bright stars in the PC chips with which to define the aperture correction,
thus making the PC aperture corrections somewhat more uncertain. This
may account for the larger differences in the comparison of the PC and
WF photometry.  Since the H~VI and H~VII fields are the two most
crowded fields in the data, the offsets and uncertainties described
here are likely larger than in the other somewhat less crowded fields.

\section{Analysis of Field Color-Magnitude Diagrams}

Each of the CMDs in Figure \ref{obs_cmds} shows evidence for stars
with a wide range of ages. The plume of stars centered around $V-I$
colors of $0.0-0.2$ is the main sequence and contains stars with ages
up to about $1-2$ Gyr old at the detection limit.  Evolved
intermediate and high mass stars undergoing core He-fusion with ages
up to $\sim 1$ Gyr are expected to populate a series of ``blue loops''
in the region between the main sequence and red giant branch. The
stellar evolutionary models predict that these stars should spend most
of their time at the red and blue ends of the loops, thus producing
two sequences in this part of the diagram. In none of the fields are
the blue loop sequences very obvious.  Each field's CMD also contains
a well-populated red giant branch (RGB) containing somewhat lower mass
stars with ages $\gtrsim 1$ Gyr that have left the main sequence but
have not yet begun fusing Helium in their cores.  After ascending the
RGB and undergoing the Helium flash at the RGB tip, stars with ages of
approximately $1-10$ Gyr end up in the red clump, the strongest
feature in each of the observed CMDs. Stars with a wide range of ages
all pile up in the red clump since the effective temperature and
luminosity are primarily determined by the mass of the He-fusing core,
which is approximately constant and independent of the initial main
sequence mass \citep{chi92}. Evolved lower-mass stars fusing Helium in
their cores that are $\gtrsim 10$ Gyr old (i.e. stars with main
sequence masses $\lesssim 1~{\rm M_{\odot}}$), would populate a
horizontal branch somewhat fainter and bluer of the red clump, as seen
for example in the CMD of Carina \citep{sme94}. However, none of the
fields show evidence for a strong horizontal branch although the main
sequence and photometric errors at the expected magnitude of the
horizontal branch would make it difficult to detect.

While all five CMDs share the same general morphology, there are
significant variations among the five fields. The average color of the
upper main sequence is bluer in the outer C25 field compared to the more
centrally located H~IV, H~VI and H~VII fields, probably indicating
differences in the average reddening for each of these fields. In
addition, the color spread of the main sequence, even at the brightest
magnitudes, appears to be larger in the bar fields compared to the C25
and H~VIII fields.  As discussed in more detail in \S4.4 below, this
may indicate some differential reddening in the bar fields. Another
difference among the five CMDs is the strength of the main sequence
relative to the red giant branch.  As a preliminary indicator of the
relative ratio of young to old stars in each of the fields, I
determined the ratio of the number of blue ($V-I<0.75$) to red
($V-I>0.75$) stars with $V<23.5$. This limit was chosen because the
photometry in all five fields is more than 90\% complete down to this
magnitude.  The ratio of blue to red stars in the the H~IV, H~VI,
H~VII, H~VIII and C25 fields is 0.52, 0.43, 0.36, 0.34 and 0.15,
respectively. These ratios are consistent with a greater recent SFR in
the bar compared to the outer regions.

There have been a number of recent studies devoted to the derivation
of the SFHs of Local Group galaxies from their CMDs
\citep{mat98}. These advances in our understanding of the evolution of
nearby galaxies are not only due to better data becoming available but
also to improved methods of analyzing the CMDs.

Deriving the ages of star clusters from observed CMDs is relatively
simple because the stars in clusters share a common distance,
extinction, age and metallicity. In these cases, simply overplotting
theoretical isochrones of the appropriate metallicity on the observed
CMD is sufficient to determine the age. The observed CMDs of field
stars in galaxies,  such as those for NGC 6822 presented in the
preceding section, are much more complicated and represent a
superposition of stars with a variety of ages and metallicities. In
some parts of the diagram, stars with different ages and metallicities
have similar colors and magnitudes, making simple isochrone fitting
inadequate to derive the ages of the stars. Recently, there have been
a number of methods proposed to extract quantitatively the star
formation history of a galaxy based upon CMDs of its stars
\citep{mar95,gal96a,gal96b,tol96,dol97,doh97,hur98,ols99,hol99,her99}.
All of these methods construct a series of model CMDs using a set of
theoretical isochrones and the results of artificial star tests.
These methods differ in many of the details such as which parameters
(i.e. distance, extinction, metal enrichment, etc.)  are assumed and
which are extracted from the CMD fit. Furthermore, these different
authors choose different statistical methods for comparing the models
and observations as well as the criteria by which the best-fitting
model is determined.  

In the remainder of this section, I describe three of these
methods that I have used to derive the field star formation history in
the five {\it HST} WFPC2 fields analyzed here. 
The method whose results
I will rely most heavily upon is that developed by \citet{dol97}
and described below in \S3.1. This method differs from the other two
described in this section in that it attempts to fit the entire CMD
all at once instead of focusing on certain phases of stellar evolution. 
Furthermore, it is able to simultaneously solve for the
star formation rate while accounting for uncertainties in other
parameters, such as the extinction and metallicity.

The motivation for choosing two other analysis methods was in part to
provide a check on the results from the more automated \citet{dol97}
approach and also to investigate the different SFH conclusions derived
from three separate analyses. Given the many different SFH analysis
methods that have been applied to different galaxies, it would be
interesting to establish whether different methods, when applied to
the same data, yield the same result.  An additional motivation for
choosing the \citet{gal96a} method was to determine whether the
results \citet{gal96a}  derived from ground-based data of NGC 6822
could be recovered from the new {\it HST} data presented here.

In contrast to the \citet{dol97} method, the methods developed by
\citet{gal96a} derive information about the SFH by comparing the
distribution of stars in the red giant branch to a series of model
CMDs constructed from a predetermined set of SFHs. By concentrating
solely on the red giant branch, this method is only able to infer
information about the SFH for ages greater than about 1 Gyr. This
method is inherently somewhat more limited than the \citet{dol97}
method because it is only possible to compare a more limited set of
SFH models with the observed CMD as described below in \S3.2 in more
detail. The \citet{gal96a} method has the advantage that it provides
easier insight into why certain models provide a better fit compared
to others. Since the \citet{gal96a} method only provides information
about the intermediate age and old SFH, the main sequence luminosity
function was also used to infer the recent SFH, as proposed by
\citet{doh97}. The results of these three analyses are presented in
\S4.

Throughout the remainder of this paper, the PC photometry is excluded
from the analysis. The CMDs of the star clusters appearing in the
PC have already been presented and analyzed in \citet{wyd00}.

\subsection{Derivation of the Star Formation History using the
\citet{dol97} Method}

In this section, I present an analysis of the SFH of NGC 6822 based
upon the methods developed by Dolphin (1997, 1999) who has kindly
provided many of the computer programs used for the following analysis.

\subsubsection{Model Ingredients}

The theoretical stellar evolutionary isochrones used here to construct
model CMDs are those calculated by the Padova group  \citep[][and the
references in both papers]{ber94,gir00}. Both sets of models follow
the evolution of stars from the zero age main sequence all the way
through to carbon ignition in higher mass stars and to the thermally
pulsing asymptotic giant branch phase in lower mass stars. They
include the most recent opacities and equations of state as well as
some amount of convective overshoot and semi-convection in stars with
convective cores.  The programs used here rely mostly on the more
recent \citet{gir00} models which describe the evolution of stars with
masses from 0.15 to 7 $M_{\odot}$ for metallicities from $Z=0.0004$ to
$Z=0.03$. For stars with masses in excess of 7 $M_{\odot}$, these
results are supplemented with the \citet{ber94} models while the
\citet{gir96} data for $Z=0.0001$ are used to extend the set of
isochrones to very low metallicities. The tracks for stars between 5
and 7 $M_{\odot}$ are very similar between the new and old sets of
isochrones \citep{gir00}, so no shift was needed to match the old and
new models. The effects of the newer ischorones on the interpretation
of stars along the red giant branch could potentially be
larger. However, the similarity between the results derived below in
the pressent work using a combination of the new and old Padua
isochrones and those from \citet{gal96a}, who relied entirely upon the
old set of models, argues that adopting one isochrone set over the
other does not lead to wildly different results.  In all of these
models, the theoretical luminosities and effective temperatures were
converted into observable $UBVRI$ magnitudes by \citet{ber94} and
\citet{gir00} using the convolution of the filter response curves with
the \citet{kur92} model stellar atmospheres.  The theoretical
isochrones are tabulated for discrete ages and metallicities.  In
order to create stars with arbitrary metallicities and ages, the
isochrones must be interpolated to any desired age and metallicity
using a set of equivalent evolutionary points along each isochrone.

In order to populate the isochrones with stars, I have assumed a power
law initial mass function of the form
\begin{equation}
\phi(\log{m}) d\log{m} = \phi_0 m^{-\alpha} d\log{m}
\end{equation} 
where $m$ is measured in $M_{\odot}$. I have assumed a constant value
of the IMF exponent of $\alpha=1.35$.
The IMF is assumed to extend from
120 $M_{\odot}$ down to 0.1 $M_{\odot}$. This leads to
an average mass of 0.352 $M_{\odot}$ and normalization of $\phi_0=0.06$.

In addition to the IMF, an extinction, reddening and distance must be
assumed in order to compare the models and observations.  In the SFH
fits described below, the extinction $A_V$ is allowed to vary while
the corresponding value of $E(V-I)$ is calculated assuming the
Galactic reddening law from \citet{odo94} with $R_V=A_V/E(B-V)=3.1$,
as tabulated for different filter systems in
\citet{sch98}. This leads to the relation $E(V-I)/A_V=0.419$.  The
ratio of $E(U-B)/E(B-V)=0.74\pm0.07$ measured by \citet{mas95} for NGC
6822 is consistent with the Galactic reddening law and supports this
assumption. Each field is assumed to have a single value of $A_V$
appropriate for all of the stars in the CMD.

Another ingredient in generating model CMDs is the chemical
enrichment law (CEL). For each model CMD, the metallicity is assumed to
increase from some initial metallicity $Z_i$ to a final metallicity
$Z_f$ following one of three chemical enrichment laws which are shown
in Figure \ref{fz} for $Z_i=0.0001$ ([Fe/H]=$-2.3$) and $Z_f=0.004$
([Fe/H]$=-0.7$). As discussed in more detail below, models
assuming various combinations
of $Z_i$, $Z_f$ and enrichment law were created and used to solve for
the SFH. These metal-enrichment laws are not motivated by any specific
physical model but rather were chosen to span a reasonable range
of possible metallicities in each time bin. A single value for the 
metallicity is assumed for each age.

\subsubsection{Description of the Method}

For a detailed description of this method, see \citet{dol97} and
\citet{dol99}. A brief overview is given in this section.

The model CMDs are constructed using the following steps. The number
of stars in the model CMDs is stored in a series of color and
magnitude bins in the CMD. The number of stars in each bin of the
model CMD will then be fit to the observed number in each bin to
determine the best-fitting SFH. A series of time bins is also chosen
for the solution with the duration of the bins generally increasing
with age to account for the fact that isochrones for older stars tend
to bunch more closely together in the CMD than the isochrones for
younger stars.

Assuming some combination of values for the extinction, metal
enrichment law, initial and final metallicities, distance and IMF, a
series of model "basis" CMDs  are created by assuming a constant SFR
of 1 $M_{\odot}$ yr$^{-1}$ for the duration of each time bin and none
at other times. A set of interpolated isochrones is then calculated
with spacing in age much smaller than the duration of the bin.  These
isochrones along with the assumed IMF are then used to determine the
number of stars formed in each CMD color-magnitude bin for this
particular time bin.  The metallicity is assumed to increase linearly
from the beginning to the end of the time bin with the initial and
final values derived from the assumed chemical enrichment law. For
each isochrone point, the artificial star test results are used to
determine in which CMD bins a star with that particular intrinsic
color and magnitude could  have been observed. The result of this
procedure is a series of model CMDs for the assumed set of input
parameters that show how stars within each time bin are distributed
within the diagram for a SFR of 1 $M_{\odot}$ yr$^{-1}$. As an
illustration of this procedure, I have created a model CMD using the
artificial star test results for the H~VIII field assuming a constant
SFR of $5\times10^{-4}$ $M_{\odot}$ yr$^{-1}$ from 15 Gyr ago to the
present, an initial and final metallicity of $[{\rm Fe/H}]=-2.3$ and
$-0.7$, respectively, and the second enrichment law shown in Figure
\ref{fz}.  The resulting model CMDs for each age bin are shown in
Figure \ref{h8tbins}.  As is evident from the figure, stars older than
about 1 Gyr are concentrated in the red clump (RC) and red giant
branch (RGB) while stars younger than a Gyr lie along the main
sequence and blue loops between the RGB and main sequence.

The choice of the time bins shown in Figure \ref{h8tbins} is somewhat
arbitrary. Ideally, one would like to chose the time bins such that
each of the basis CMDs is significantly different from one-another to
avoid degeneracies in the fitting while at the same time being able to
reproduce the full range of possible SFHs. While such an investigation
is beyond the scope of this paper, it is possible to test whether the
time bins shown here significantly bias the SFH solutions derived.
To test this, I have solved for the SFH in the H~VIII field using the
fitting procedure and assumptions described below using three
different choices for the time bins. For the first test, 12 time bins
were selected from 10 Myr to 15 Gyr, varying in length from 0.2 Gyr
for the youngest ages to 2 Gyr for the oldest bin. For the second and
third tests, the time bins were chosen with duration  $\Delta {\rm
log}(t)=0.1$ and  $\Delta {\rm log}(t)=0.2$, respectively, between 10
Myr and 15 Gyr. In all three tests, the solutions were determined
using the assumptions about the metallicity, extinction, distance,
etc. described below. The resulting SFRs vs. time are plotted in
Figure \ref{timeres}. The three panels in the first column show
the solutions at their full resolution while the second column shows
the same solutions rebinned to match as closely as possible
the time bins used for the final results presented below in \S4.
Despite the varying number of time bins in each solution, the
resulting SFRs, when binned to the same resolution are similar to
within the errors.

The object of the fitting procedure is then to determine which linear
combination of these single time bin CMDs best represents the data.
More specifically, the number of model stars present in the $i$th
color-magnitude bin of the CMD, $N_i$, is given by
\begin{equation}
N_i = \sum_j{c_j M_{ij}}
\end{equation}
where $c_j$ is the coefficient for the $j$th time bin and
$M_{ij}$ is the number of stars in the $i$th CMD bin of the
$j$th time bin assuming a SFR of 1 $M_{\odot}$ yr$^{-1}$. Thus,
the $c_j$ coefficients are the SFRs (in units of $M_{\odot}$ yr$^{-1}$)
for each time interval.

In order to determine the SFH that best represents the data, a parameter
is needed to describe how well a model fits the data. Ideally,
one would want to use a $\chi^2$ statistic for the comparison. However,
in practice bad points in the CMD, foreground stars and other objects
which contaminate the CMD bias such a statistic making it difficult
to derive a good fit. In addition there are certain evolutionary phases,
particularly the blue loops and red clump where the models are
more uncertain. To avoid having the fitting statistic being biased
towards areas of the CMD where
there are observed stars but few or no model stars, a modified $\chi^2$
statistic was adopted which I will denote as $\bar{\chi}^2$. It is given by
\begin{equation}
\bar{\chi}^2 = \sum_i{\frac{(N_i - n_i)^2}{N_i}}
\end{equation}
where $n_i$ is the number of observed stars in the $i$th CMD bin and
$N_i$ is the number of model stars.  In order to account for poorly
fit stars as well as some amount of foreground contamination, a term
$xLF$ is added to the value of $N_i$ in each bin, where $LF$ is the
number of model stars in that interval of magnitude, regardless of
color, and $x$ varies linearly from 0.1 at the bright end of the CMD
to 0.4 at the faint end. In addition, if in any particular bin
$\bar{\chi}_i^2>9$, then the value for that bin is replaced by the
value of $(6\bar{\chi}_i-9)$ so that the value of $\bar{\chi}^2$ does
not blow up so quickly for large differences between the model and
data. This particular modification was chosen such that both
$\bar{\chi}^2$ and its derivative are continuous.  The set of SFR
coefficients producing the minimum value of  $\bar{\chi}^2$ is then
the best-fitting solution. Unfortunately, given the modifications to
the fit parameter from a traditional $\chi^2$, it is difficult to
quantify the significance of the solutions. A detailed investigation
of the statistical significance of an individual SFH solution is
beyond the scope of the present work.  Nevertheless, the value of
$\bar{\chi}^2$ should rank the solutions in a relative sense.

The above discussion of fitting the observed CMD describes a solution
in which only the SFR(t) was allowed to vary while all of the other
input parameters (extinction, metal enrichment law, initial metallicity,
final metallicity, IMF, distance) were assumed constant. Given good
enough photometry, the effects of these parameters on the CMD are
different enough so that each of these parameters can be allowed to
vary and solved for as a parameter in the fit \citep{dol97}.  In such a
case, the procedure described above is repeated many times with each
of the parameters allowed to vary within some specified range.  Then
some number of the best-fitting solutions are chosen. A weighted
average of these solutions then provides estimates of all of the input
parameters as well as the SFH. Furthermore, the spread in the SFRs and
the other parameters can be used as an estimate of their probable
errors. The data presented here do not go far enough down the main
sequence to make such a full solution possible.  Since some of these
parameters are fairly well constrained by other observations of NGC
6822, I have only allowed some of these parameters to vary while fixing
others.

A summary of the parameters that I have fixed as well as the ranges
tried for other parameters is shown in Table \ref{sfhpar}.
As discussed in the
introduction, the distance to NGC 6822 is fairly well-established. In
all of the fits, I have assumed the Cepheid distance modulus of
$(m-M)_0$=23.49 as derived by \citet{gal96c}.  The extinction $A_V$
was allowed to vary from 0.6 to 1.3 mag in steps of 0.1, thus
bracketing the range of extinctions (including both foreground and
internal) that have been estimated for NGC 6822 in the past
\citep{mas95}. 

The current oxygen abundance in the gas of NGC 6822 is substantially
less than solar, as first shown by \citet{pei70}. \citet{pag80}
measured the oxygen abundances for seven \ion{H}{2} regions and found
an average abundance of $12+{\rm log}(O/H)= 8.25\pm0.07$ which
corresponds to [Fe/H]$=-0.7$ for a solar O/Fe ratio. Confirming these
results, \citet{ski89b} determined an abundance for the \ion{H}{2}
region H~V of $12+{\rm log}(O/H) = 8.20$. In contrast to those
studies, the average abundance of the six \ion{H}{2} regions observed
by \citet{cha00} is $12+{\rm log}(O/H) = 7.91\pm0.06$, corresponding
to an overall metallicity of [Fe/H]$=-1.0$. Unfortunately, there are
no \ion{H}{2} regions in common between the \citet{cha00} and
\citet{pag80} samples, so it remains unclear whether the different
results represent a real abundance spread or whether there is some
zeropoint difference between the two studies.  In addition to the
\ion{H}{2} regions, abundances for individual high mass stars in NGC
6822 are beginning to be determined. Based upon high resolution
spectroscopy, \citet{ven01} determined the oxygen abundance for two
A-type supergiants to be similar to the \citet{pag80} results. The
mean oxygen abundance for these two stars is $12+{\rm
log}(O/H)=8.36\pm0.19(\pm0.21)$ where the two uncertainties 
correspond to random and systematic errors, respectively.
Furthermore, the iron lines in the stellar spectra lead to a solar
O/Fe ratio: $[{\rm O/Fe}]=+0.02\pm0.20(\pm0.21)$. These value stand in
contrast to observations of Galactic low metallicity stars which show
an overabundance of oxygen relative to other heavy elements
\citep{whe89}. The \citet{ven01} results are consistent with spectra
of two B-type supergiants that have an average metallicity of $[{\rm
Fe/H}]=-0.5\pm0.2$ \citep{mus99}. The different metallicities
determined from these various studies could indicate a possible spread
in the current metallicity of NGC 6822 although no trend with
galactocentric distance is yet obvious. Due to these lingering
uncertainties, I have allowed the current metallicity to vary from
$[{\rm Fe/H}]_f=-0.7$ to $-1.1$ in steps of 0.2 dex.

The only observational constraints on the initial metallicity in NGC
6822 that are independent of the distribution of stars on the RGB
come from observations of the cluster H~VII. Based upon its
integrated spectrum, \citet{coh98} derived a metallicity of
[Fe/H]$=-1.95\pm0.15$ dex and age of $11^{+4}_{-3}$ Gyr. These results
are in general agreement with the age of $(10\pm2)$ Gyr and
metallicity of $[{\rm Fe/H}] \sim -2.0$ determined by \citet{cha00}.  The CMD
and structure of H~VII from the {\it HST} data also support the
conclusion that H~VII is likely similar to Milky Way globular
clusters \citep{wyd00}. The initial metallicity for the SFH solutions
was chosen to bracket the H~VII metallicity and was allowed to vary
from [Fe/H]$_i=-2.3$, the lowest value available from the stellar
evolutionary models, to [Fe/H]$_i=-1.4$ in steps of 0.3 dex.

When fitting their ground-based CMDs of NGC 6822,
\citet{gal96a} found that a common feature of accepted models
was that stars in the $6-9$ Gyr age range have metallicities
of $Z \approx 0.002-0.003$, or $[{\rm Fe/H}] \approx -0.9$.
The accepted models either had star formation beginning 12-15 Gyr
ago at a low metallicity ($Z_i=0.0001$) or star formation beginning
$6-9$ Gyr ago from gas at a somewhat higher metallicity ($Z_i=0.002$).
In all of their models a metal enrichment law linear in $Z$ was assumed
leading to a metallicity of $Z=0.002-0.003$ for stars formed $6-9$
Gyr ago. This metallicity for $6-9$ Gyr-old stars is roughly
consistent with the first two chemical enrichment laws shown in 
Figure \ref{fz}, regardless of the initial or final metallicity.
However, it is inconsistent with the chemical enrichment law that
is linear in [Fe/H] for low initial metallicities. As is shown in the
next section, these models tend to provide a worse fit to the data
than the other enrichment laws.

With these choices of parameters, all of the SFH solutions were
performed using the following steps.  For each solution the chemical
enrichment history was fixed by choosing one of the possible
combinations of [Fe/H]$_i$ and [Fe/H]$_f$ as well as one of the three
enrichment laws shown in Figure \ref{fz}. Then a solution was
determined for a series of values of $A_V$, beginning with
$A_V=0.9$. Next the value of $A_V$ was increased or decreased in steps
of 0.1 mag until the minimum of the fit parameter $\bar{\chi}^2$ was
found. The value of $A_V$ was constrained to lie within the range
specified in Table \ref{sfhpar}.  This was repeated for each
combination of [Fe/H]$_i$, [Fe/H]$_f$ and chemical enrichment law. The
final values of the SFR and metallicity for each time bin and the
best-fitting value of $A_V$ were determined by averaging together some
number of the best-fitting solutions. The spread in each of the parameters
as well as the SFRs in each bin among the best solutions provides
an estimate of the likely uncertainties in each of the parameters.

\subsection{Derivation of the Star Formation History using
the \citet{gal96a} Method}

A somewhat different approach in deriving the SFH has been used by
other authors. These approaches have in common that they concentrate
on certain specific regions of the CMD for comparison of model diagrams
with the observed data. Although such an approach has the limitation
that it does not make use of all of the information in the CMD, it
does have the advantage of allowing a better understanding of why
certain models provide a better fit compared to others. In this
section, I attempt to duplicate the methods used by
\citet{gal96a} to derive the SFH for the old and intermediate age
stars for the H~VIII field. 

\subsubsection{Model Ingredients}

Many of the ingredients used to construct the models used
in \S{3.1} are also used here. The evolutionary models
are the same combination of the \citet{gir00} and \citet{ber94}
results described above. The distance and IMF are identical to those
assumed already as well. For the H~VIII field, I have assumed 
an extinction of $A_V=0.82\pm0.1$, the value determined by the 
full SFH solutions via the \citet{dol97} method. (See \S4.2).

The same series of SFH shapes adopted by \citet{gal96a} were assumed
here and are shown in Figure \ref{sfhshapes}. The main goal of adopting these
shapes is to determine the overall trend in the SFR with time.  For
each SFH shape, three equal length time bins are assumed from the time
of initial star formation $T_i$ to the present. In practice, the
recent star formation is truncated at 10 Myr. The SFR is allowed to
change amplitude once, either at $1/3T_i$ or at $2/3T_i$. The higher
SFR is assumed always to be four times the lower SFR. In addition, a
constant SFR case is considered. All five shapes are normalized to
the same integrated mass of stars formed. The SFH shapes have been
ordered such that they progress from a predominantly young population in
Shape 1 to a predominantly older population in Shape 5.
Shape 3 is the constant SFR case. Four values of $T_i$ have been
considered: 15, 12, 9 and 6 Gyr.

The same metal enrichment law assumed by \citet{gal96b} is assumed
here. The metallicity is assumed to increase linearly in $Z$ from its
initial value $Z_i=0.0001$ ([Fe/H]$=-2.3$)
to the current metallicity of $Z_f=0.004$
([Fe/H]$=-0.7$). Models with the same final metallicity but with $Z_i=0.002$
were also generated.

For each value of $T_i$, a synthetic CMD was produced
assuming a constant star formation rate in the three time bins of
$5\times10^{-3}$ $M_{\odot}$ yr$^{-1}$, a rate about 10 times that in
the observed H~VIII CMD. Thus, many more model stars are created than
observed stars to reduce the noise in the model CMD. Then the number
of stars created in each time bin was scaled relative to one
another to create model CMDs for each of the five SFH shapes. The
observational errors and incompleteness were incorporated in the models
using the results of the artificial star tests. There are therefore
a total of 40 model CMDs given all the possible combinations of the
four values of $T_i$, the five SFH shapes and the two values of $Z_i$.

\subsubsection{Definition of Indicators Relative to the Red Giant Branch}

In order to compare their model and observed CMDs, \citet{gal96a}
calculated several indicators relative to the distribution of stars
in the ``red tangle,'' a region containing the RGB, and the
``red tail,'' the region of the CMD above and redward of the
RGB consisting primarily of asymptotic giant branch (AGB)
stars. In particular the color
distribution of stars in the red tangle in the magnitude interval
$-3.75<M_I<-3.25$ was used to discriminate among the different
models.

The CMDs analyzed here do contain a few stars brighter than the RGB
which are likely part of the red tail structure seen in the
ground-based CMDs. It is difficult to use these stars to place
constraints on the SFH due to their small numbers.  In addition, the
saturation limit in the {\it HST} data occurs at $I\approx19$, a level
only about 1 mag above the TRGB. Thus, I have chosen to use the
distribution of stars along the RGB and RC to differentiate between
the different models. The definitions of the regions in the CMD used
for the analysis are shown in Figure \ref{cmdregions} superposed on
the $M_I, (V-I)_0$ CMD for the H~VIII field. There are four regions
along the RGB as well as a region containing the RC. The brightest RGB
region was chosen to be nearly identical to the region of the red
tangle used by \citet{gal96a}.  Similar to the AGB, the RC contains
intermediate age stars $1-10$ Gyr old. However, it is less sensitive
to age and metallicity as shown below in \S4.

For each of the 40 model CMDs, the average SFR was calculated by
scaling the number of model stars in the four RGB regions to match the
observations. This was accomplished by minimizing the $\chi^2$
statistic given by
\begin{equation}
\chi^2 = \sum_{i=1}^4{\frac{(N_i-n_i)^2}{N_i}}
\end{equation}
where $N_i$ is the number of model stars in the RGB bin and $n_i$ is
the number of observed stars in the same bin.

For each combination of $T_i$, SFH shape and $Z_i$ the color
distribution of stars in the brightest RGB region was calculated,
scaled to the observations using the average SFR for that model. A
grid of the resulting model color distributions for the brightest RGB
region are shown as the solid lines in Figure \ref{rgbcdist} for the
$Z_i=0.0001$ case.  The observed color distribution is plotted in each
panel as the dotted line. Each row of the figure corresponds to a SFH
shape which is labeled in the corner of every panel while each column
corresponds to the value of $T_i$ labeled at the top.  It is clear
that the models with star formation beginning as soon as 6 or 9 Gyr
ago tend to produce RGBs that lie blueward of the observed RGB. For
each value of $T_i$, a secondary blue peak in the RGB color
distributions for SFH shapes 1, 2 and 3 appears. The peak becomes more
prominent relative to the red peak as the proportion of young stars
increases. The stars in the blue peak are intermediate mass evolved
stars with ages $< 1$ Gyr that lie along the red end of the blue loops
in the CMD. The lack of this secondary peak in the observed CMD argues
against a SFH with most of the star formation occurring within the
most recent third of the galaxy's history.

To better quantify the differences between the model and  observed
color distributions, I have defined three indicators all of which
were used by \citet{gal96a} in their analysis. The first
is the median $V-I$ color of the RGB. The second indicator is
C95 which marks the red edge of the RGB. It is calculated by finding
the color below which 95\% of the stars lie. The third is a $\chi^2$
statistic similar to that shown in equation 4 except that now
the summation is made over bins in color. I have restricted the
calculation to only include those color bins where the number of model
stars is non-zero and the number of observed stars is greater than six.
The latter limit was chosen to insure that the results are not dominated
by the bins with few stars.

In addition to the color distribution of stars in the RGB, I have
calculated for each model the ratio of stars in the RC region to the
number stars in each of the RGB regions.  When stars reach the RC and
begin core He fusion, the mass of the He-burning core is similar for
stars with a wide range of initial masses \citep{chi92}. This means
that the amount of time a star spends in the RC is relatively
insensitive to its original main sequence mass. On the other hand, the
time it takes for a star to ascend the RGB does tend to increase with
decreasing mass \citep{chi92}. As a consequence, one would expect the
ratio of RGB stars to RC stars to increase with the dominant age of
the stellar population. For a more detailed discussion concerning the
use of this ratio in constraining the SFHs of Local Group dwarf
galaxies see \citet{tol98} and \citet{col99}.

\subsection{The Recent Star Formation History from the Main Sequence
Luminosity Function}

While the old and intermediate age stars are concentrated in the
RGB and RC structures in the CMD, the younger stars with ages less
than about 1 Gyr, are distributed along the main sequence (MS) and
He-burning blue loops. Along these sequences, stars
of similar masses have similar brightnesses, a fact which can be exploited
to determine the recent SFH from the luminosity functions of stars
in these two parts of the diagram. \citet{doh97} developed a method
to derive the SFH from the MS luminosity function which I have applied
here to all five {\it HST} fields in NGC 6822.

Although stars have similar masses for each magnitude along the main
sequence, they can have any age less than or equal to the MS turn-off
age for that mass.  The \citet{doh97} method relies on the following
equation for the number of stars $C_n(M_V)$ within the $n$th magnitude
bin of the MS luminosity function:
\begin{equation}
C_n(M_V) =  \int_{m_1}^{m_2} \phi(m) dm \sum_{i=1}^{n}{\frac{R_i}{<m>}
\Delta t_i}
\end{equation}
where $R_i$ is the SFR in the $i$th bin in units of $M_{\odot}$
yr$^{-1}$, $<m>$ is the average mass (in $M_{\odot}$) weighted by the
IMF and $\Delta t_i$ is the difference in turnoff age between the
beginning and end of the magnitude bin.  $\phi$ is the IMF defined by
equation 1, with the same exponent and normalization as used for the
automated fits of the CMD. Finally, $m_1$ and $m_2$ are the
corresponding MS turnoff masses at the faint and bright ends of the
magnitude bin, respectively. These equations can be solved for the
values of $R_n$ by taking the difference between neighboring bins
with the convention that $n$ increases with magnitude.
Thus the value of the SFR corresponding to the $n$th magnitude bin is
given by
\begin{equation}
R_n = \frac{<m>}{\Delta t_n} \left(\frac{C_n}{A_n} - 
\frac{C_{n-1}}{A_{n-1}}\right)
\end{equation}
where $A_n$ is the integral over the IMF between the minimum and
maximum turnoff masses for the $n$th magnitude interval. The minimum
$V$ magnitude in each CMD was limited to 20 due to saturation. The SFR
corresponding to  brighter magnitudes was set to zero.

To apply these equations to the data the MS turnoff age and the MS
turnoff mass are needed as a function of $M_V$. These relations were
taken from the \citet{ber94} stellar models.
For each field the model data were interpolated to the
endpoints of each of the magnitude bins used to construct the main
sequence luminosity function. The conversion between
absolute and apparent magnitude was calculated assuming a distance
modulus of 23.49 and the extinctions listed in Table \ref{sfhtab1}
which were derived using the methods presented in \S3.1.

The curve used to define the main sequence is shown in Figure
\ref{cmdregions} as the dashed line superposed on the H~VIII CMD. The
main sequence stars were defined to be all stars that lie blueward of
this line. For the other fields, the curve was shifted according to
the appropriate reddening for each field. The observed MS luminosity
function was then calculated for each field using 0.5 mag wide bins
down to a limiting apparent magnitude of $V=25.5$ where the
completeness in all five fields is greater than 30\%. A set of
artificial stars with input $V-I$ colors greater than 0.0 and less
than the same curve used to define the observed MS was chosen to
determine the completeness as a function of magnitude for the
MS. Finally, the recent SFHs for each field were calculated from the
completeness corrected luminosity functions using equation 6. The
errors in each of the SFRs were determined from the errors in the
completeness corrections as well as the Poisson errors in the number
of observed stars in each bin. The Poisson errors dominate the
total error in the SFR for each bin.

Before proceeding with presenting the results, it is important to
emphasize some of the limitations common to all three methods
presented in this section.  First, there is undoubtedly some fraction
of unresolved binary stars in the CMDs whereas the SFH solutions used
here assume that it is zero.  Second, there could be some intrinsic
spread in metallicity at a given age which the solutions given here
assume is zero. Third, I have neglected any differential reddening
within each field. Fourth, I have assumed a single power law IMF with
exponent equal to the original value determined by
\citet{sal55}. While there is some evidence that the form of the IMF
does not vary widely with environment, measurements of the IMF are
uncertain enough that significant variations with time or among
different galaxies can not be ruled out \citep[e.g.][]{ken98b}.
Finally, and probably most importantly, these solutions ignore any
possible systematic errors and uncertainties in the stellar
evolutionary models. As was noted by \citet{hol99}, all of the
derivations of the SFHs of Local Group galaxies produce model CMDs
which are statistically incompatible with the observations. This is
likely a result of the many approximations used in deriving the
best-fitting SFH but is also likely in part due to the uncertainty in
the stellar models.

\section{Results and Discussion}

In the previous section, I have described the different methods
utilized to infer information about the star formation history of
NGC 6822 from the observed CMDs. In this section, the results of
these analyses are presented.

\subsection{Age of the Oldest Stars}

One of the most basic questions we would like to know about galaxies
is when their first stars began forming.  The cluster H~VII has a
likely age of $11^{+4}_{-3}$ Gyr, indicating that star formation began
at least about 8 Gyr ago \citep{coh98}.  On the other hand,
\citet{gal96a} concluded that star formation in NGC 6822 most likely
began about 15 Gyr ago based upon their ground-based CMDs. They also
showed that the CMD could also be consistent with star formation
beginning as soon as 6 Gyr ago if there were prompt metal enrichment
at the onset of star formation to half its current value. The
metallicity of H~VII measured by \citet{coh98} would argue against
this possibility although it is possible that H~VII and the field
stars formed during the same period may not have shared the same
metallicity.

Using the techniques developed by \citet{dol97} and outlined
in \S3.1, I have explored to what extent the {\it HST} data can be used
to place constraints on the time when star formation began in NGC 6822.
For this experiment, I have concentrated on the H~VIII field due to
its somewhat lower crowding than the other Hubble cluster fields.
Using the parameters listed in Table \ref{sfhpar},
I have solved for the SFH in
this field under three different assumptions about the ages of the
oldest stars, $T_i$. In each solution, the same eight time bins were
used except for the beginning of the oldest time bin. I have chosen to
perform the solutions for $T_i=15$, 12 and 9 Gyr. 

The results for a subset of the solutions are plotted in Figure
\ref{h8tiz} for an assumed extinction of $A_V=0.8$.  Each panel of the
figure plots the value of the fit parameter $\bar{\chi}^2$ as a
function of $T_i$ for a particular combination of $[{\rm Fe/H}]_i$ and
$[{\rm Fe/H}]_f$. The solid, dotted and dashed lines in each panel
correspond to the enrichment laws 1, 2 and 3, respectively, as plotted
in Figure \ref{fz}. For clarity only the solutions for the extreme
values of $[{\rm Fe/H}]_i$ are shown here.

While in general the value of the fit parameter decreases with
increasing $T_i$ for a given chemical enrichment history, this
conclusion depends sensitively on the assumptions made about the metal
enrichment history.  For a given value of the initial metallicity,
increasing the final metallicity tends to decrease the differences in
fit parameter among the three values of $T_i$. This somewhat
counterintuitive result is due to the fact that changing the value of
$[{\rm Fe/H}]_f$, while holding the enrichment law and $[{\rm
Fe/H}]_i$ constant, increases the spread in metallicity in the oldest
time bin, thus allowing the CMD to be fit equally well by a SFH with a
smaller range in age. The chemical enrichment law 1, the chemical
enrichment law with the most rapid increase in metallicity with time,
results in the best fit for a specific choice of initial and final
metallicity, particularly for low values of the initial
metallicity.  Increasing the initial metallicity tends to decrease the
differences in fit parameter among the three different enrichment laws
because decreasing the difference in initial and final metallicities
tends to make the chemical enrichment histories derived from the three
chemical enrichment laws more similar to one another.

For each value of $T_i$, an average SFH and metal enrichment history
was derived by averaging together the approximate 30 best solutions.
The results are summarized in Table \ref{h8tab1} which lists the
minimum fit parameter found for each choice of $T_i$  as well as the
maximum fit parameter used to select the best solutions. The third
column lists the resulting number of solutions averaged.  The derived
value of $A_V$ appears to increase as $T_i$ decreases although the
difference is small. While one should keep in mind the systematic
trends with metallicity shown in Figure \ref{h8tiz}, overall the fits
tend to worsen as $T_i$ decreases.  Table \ref{h8tab2} lists the SFRs
as a function of age for each value of $T_i$ while Figure
\ref{h8tisfh} plots the SFRs and metallicities.  As is evident from
the tables, the star formation rate for ages up to 3 Gyr is relatively
unaffected by the choice of $T_i$. For older ages, however, the SFR in
the oldest time bin increases as $T_i$ decreases. Furthermore, the
star formation rate in the $5-7$ Gyr bin decreases with $T_i$ while
the star formation rate in the $3-5$ Gyr bin increases.

The best-fitting average SFRs and metallicities plotted in Figure
\ref{h8tisfh} were used to construct model CMDs which are shown in
Figure \ref{h8timodel} for each value of $T_i$ along with the observed
CMD. As would be expected, the main difference in the morphology of
the CMD is the distribution of stars along the RGB. As $T_i$
decreases, the RGB tends to become narrower. As is well known, there
is an age-metallicity degeneracy for RGB stars in that a blue RGB can
be produced by either an old, metal-poor population or by a younger
and more metal-rich one. On the other hand, the width of the RGB
could be increased somewhat if a spread in metallicity at each age
were allowed.

The results of the analysis using the \citet{gal96a} methods, as
described in \S3.2, point toward a similar conclusion regarding the
age of the oldest stars in NGC 6822.  Plots of the median color, C95
color and $1/\chi^2$ as a function of SFH shape are shown in Figure
\ref{cmdvar1} for the brightest RGB region 1.  The solid lines in each
plot show the observed value and, in the case of the median and C95
colors, the $\pm1\sigma$ error bars, calculated assuming a $\pm0.05$
uncertainty in the reddening.  While the median and C95 colors are
rather insensitive to the SFH shape, they are able to discriminate
between different values of $T_i$. Consistent with expectations, the
RGB tends to become bluer as the time of initial star formation
decreases. The differences among the different models are largest for
the brightest RGB bin and become more similar to each other for the
fainter bins.  Thus only the results for the brightest bin are
considered here.  The only model whose indicators are all consistent
with the  observed values is the $T_i=15$ Gyr case with the $T_i=12$
Gyr model being just barely excluded by the C95 indicator.  This same
conclusion is supported by the values of $1/\chi^2$.  The models with
star formation beginning 12-15 Gyr ago and SFH shapes 3 and 4 provide
the best fits. It is important to bear in mind that these conclusions
rest on the assumption of a low initial metallicity. Similar fits were
attempted for an initial metallicity of $[{\rm Fe/H}]=-1.0$ and the
results are plotted in Figure \ref{cmdvar2}.  For this higher
metallicity, the median and C95 colors are consistent with values of
$T_i$ from $6-12$ Gyr and show a slight preference for SFH shapes 1
and 2. The value of $1/\chi^2$ for all the models is less sensitive to
the SFH shape and do not reach as high values as for the low
metallicity case.  These results are entirely consistent with the
conclusions reached by \citet{gal96a} based on the same analysis of
their ground-based CMDs in that the distribution of stars near the tip
of the RGB can be fit by star formation beginning at younger ages if
the initial metallicity is high.

In Figure \ref{cmdvar1}(c), the ratio of the number of stars in RGB
region 1 to the number of stars in the RC is shown as a function of
SFH shape. The model ratios are very similar for all values of $T_i$
and are not able to reproduce the observed ratio.  For the models
plotted in \ref{cmdvar1}(c), there is a small increase in the ratio
from SFH shapes 1 to 5, consistent with the rising fraction of older
stars from Shapes 1 to 5. The origin of the difference is not obvious
and may be due to some combination of uncertainties in the stellar
models, the limited set of SFH solutions attempted here or possibly a
non-Salpeter IMF. In particular, the He-burning lifetimes of stars in
the RC depend strongly upon the assumptions made concerning covective
overshoot which may lead to systematic uncertainties in the stellar
models \citep[e.g.][]{ber01}.  The same discrepancy is apparent in the
high initial metallicity fits in Figure \ref{cmdvar2}. Although not
plotted in these figures, the RGB to RC ratio in the models is lower
than the observations for each of the RGB regions shown in  Figure
\ref{cmdregions} individually. Thus, the discrepancy would not be
solved by including stars further down the RGB in the calculation.

In summary, both the \citet{dol97} and \citet{gal96a} analysis are best
able to fit the observed CMD with star formation beginning 15 Gyr
ago. However, it is important to emphasize that  depending upon the
assumptions made about the metal enrichment history, the CMDs can also
be fit by star formation beginning at younger ages. In particular, the
trends in Figure \ref{h8tiz} show that the observed CMD of NGC 6822
can be fit equally well by star formation beginning only 9 Gyr ago,
even for low initial metallicities, if the metallicity increase with
time is rapid. This is similar to the conclusion reached by
\citet{gal96a} that assuming a high initial metallicity allows the RGB
to be fit by younger ages. More definitive conclusions about when NGC
6822 began forming stars must await either deeper CMDs or independent
information about the abundances of stars along the RGB.

One of the hallmarks of a truly old, globular-cluster age population
($\gtrsim 10$ Gyr old) is the presence of a horizontal branch.  None
of the CMDs presented in the preceding chapter show any evidence for a
strong horizontal branch, which would be expected somewhat fainter and
blueward of the red clump as seen, for example, in the CMD of Carina
\citep{sme94}. Still, the photometric errors at the expected level of
the horizontal branch around $V\approx25$ as well as the strong main
sequence may be masking any horizontal branch, if present. No searches
for RR Lyrae variables, which are readily identifiable horizontal
branch stars, have yet been published for NGC 6822.

\subsection{Old and Intermediate Age 
Star Formation Histories for all Five Fields}

The SFHs for the five {\it HST} fields were derived using  the
\citet{dol97} method with the parameters listed in Table \ref{sfhpar}
and the same time bins as for the H~VIII solutions already
presented. As $T_i=15$ Gyr provides the best fit to the H~VIII data, I
have adopted this value for all of the fields.  Before discussing the
results for each of the fields, I will first explore the systematic
effects of the choice of extinction and metallicity on the derived
SFHs using the results for the H~VIII field.

The effects of the value of the extinction on the derived SFH are
shown in Figure \ref{h8avsfh} for the H~VIII field in which the SFR
and metallicity are plotted as a function of age assuming fixed values
for the extinction
of $A_V=0.6$, 0.9 and 1.2 in the top, middle and bottom panels,
respectively. As $A_V$ increases, the best-fitting value of the
current metallicity tends to decrease while leaving the initial metallicity
unaffected. Holding the metallicity fixed and increasing the value of
$A_V$ (and the corresponding reddening)  has the effect of making the
main sequence redder while for a fixed value of the extinction,
increasing the metallicity tends to make the main sequence bluer.
Therefore, the SFH-fitting program attempts to compensate for the
reddening of the main sequence for higher values of the extinction by
decreasing the best-fitting current metallicity.  In addition,
increasing the extinction tends to increase the recent SFR as well as
the ratio of young to old stars. However, there is a clear minimum in
the value of the fit parameter for values of $A_V=0.7-0.9$ for this
field, as shown in Figure \ref{h8fit}(a), where the fit parameter
is plotted as a function of $A_V$.

In general the CMDs presented here do not go deep enough to place
strong constraints on the age-metallicity relation in NGC 6822.  The
value of the fit parameter resulting from fitting the H~VIII field CMD
is shown in panels (b), (c) and (d) of Figure \ref{h8fit} as a
function of the initial metallicity, final metallicity and chemical
enrichment law, respectively. The value of the initial metallicity is
not very well constrained by these CMDs. This is not entirely
unexpected since all of the information about the SFH for ages $> 1$ Gyr
is derived from the RGB and RC, areas of the CMD  where the age and
metallicity tend to be degenerate. There is a slight tendency for the
fits to favor higher values of the final metallicity.  The chemical
enrichment laws 1 and 2 are somewhat favored over law 3 although the
dependence of the fit parameter on the chemical enrichment law is
slight.

The choice of age-metallicity relation does systematically affect the
derived SFRs.  The effects of the choice of initial metallicity and
chemical enrichment law on the SFRs is illustrated in Figures \ref{h8amr1} and
\ref{h8amr2}.  In each case, the final metallicity was held constant
at a value of $[{\rm Fe/H}]_f=-0.9$ while the initial metallicity was
assumed to be $[{\rm Fe/H}]_i=-2.3$ in Figure \ref{h8amr1} and $-1.4$
for those solutions shown in Figure \ref{h8amr2}.  In each figure, the
SFR is shown on the left-hand side while the assumed age-metallicity
relation is shown on the right. For low initial metallicities, the
ratio of young to old stars tends to increase going from chemical
enrichment laws 1 to 3, corresponding to a less rapid increase in
metallicity with time. Furthermore, the SFR in the $7-15$ Gyr time bin
decreases by a factor of $\sim3-4$ as the initial metallicity goes from
$-2.3$ to $-1.4$. As would be expected, the three SFHs shown in
Figure \ref{h8amr2} for a higher initial metallicity are much more
similar to one another. The smaller difference in initial and final
metallicities means that the choice of chemical enrichment law makes
less of a difference in the resulting age-metallicity relation.

Keeping these systematic effects in mind, the approach adopted here is
to average together the SFRs from some number of the solutions with
the lowest value of the fit parameter.  Unfortunately, as already
discussed above, the only constraints presently available on the
age-metallicity relation in NGC 6822 are the value of the current
metallicity, as measured from the \ion{H}{2} regions and individual
massive stars, and the metallicity of the old cluster H~VII. By
averaging together the results from solutions resulting from different
assumptions about the initial and final metallicities as well as the
enrichment law, I may not necessarily obtain a more accurate
result. However, the spread in SFRs resulting from these solutions
allows an estimate to be made of the uncertainty in the SFR resulting
from our ignorance of the chemical enrichment history in NGC
6822. Clearly, additional observations constraining the
age-metallicity relation in this galaxy are needed to better
understand its SFH.

As before, approximately 30 of the best solutions per field were used
to calculate the average  and standard deviation of the SFRs,
metallicities and extinctions. The minimum and maximum fit parameters
used as well as the best-fit value of the extinction for each field
are shown in Table \ref{sfhtab1} while the corresponding SFRs for each
field are listed in Table \ref{sfhtab2}.  The SFHs and corresponding
metal enrichment histories  for all the fields are plotted in Figures
\ref{sfh1} and \ref{sfh2}.

The most striking property of the SFHs in Figures \ref{sfh1} and
\ref{sfh2} is the similarity in the SFH shapes for all five
fields. The SFR appears to be constant or perhaps slightly increasing
from $15-5$ Gyr ago. Then after a decrease in the SFR from 3 to 5 Gyr
ago, it rises back up to about the same level again in all the fields
in the $0.6-3.0$ Gyr age bin. The overall normalization of the SFR is
somewhat higher in the H~VI and H~VII fields that lie in the bar than
in the H~IV, H~VIII and C25 fields.  The different average SFRs
roughly correlate with the surface density of stars in each
field. Thus, the bar contains a higher density of the old and
intermediate age stars in addition to a higher density of younger
stars.

The SFRs in the H~VI and H~VII fields, which have substantial overlap,
agree to within the errors for stars older than 0.6 Gyr.  The
differences between the two fields are some indication of the internal
consistency of the results presented here. For the two most recent
time bins, the SFR in the H~VI field is somewhat higher. The number of
stars with $V-I<0.7$ and $V<23.5$ is 1902 in the H~VI field and 1550
in the H~VII field, a difference of about $20\%$. The higher recent SFR
in the H~VI field is thus the result of the higher density of bright
blue stars compared to the H~VII field.

It should be pointed out that the time bins chosen for this analysis
are somewhat arbitrary. In particular, as is evident from Figure
\ref{h8tbins}, the CMDs for stars with ages between about 1-7 Gyr are
quite similar to one another. This means that the dip in the SFR in
all fields 3-5 Gyr ago may not be very  significant. Hence, the main
conclusion to be drawn from the SFHs in Figures \ref{sfh1} and
\ref{sfh2} is that overall the SFR in NGC 6822 has been fairly
constant with no evidence for strong variations with time.

These results derived using the full fits to the CMD from
the \citet{dol97} method are generally consistent with the old
and intermediate age star formation history derived using the
\citet{gal96a} method. The best-fitting model is the one in which 
the star formation rate is either constant or shows a factor
of four decrease in the last 5 Gyr compared to the older SFR.
These two SFHs are plotted in Figure \ref{h8sfh} along with the
results for this field from Figure \ref{sfh2}. While the SFHs
derived from the two different methods are roughly consistent,
the SFR from 5$-$15 Gyr ago is somewhat higher for SFH shape
4 than in the fit from the \citet{dol97} method. This is likely a
consequence of the requirement that the SFR is only allowed to decrease
by a fixed factor of four.

\subsection{Recent Star Formation Histories}

The recent SFRs from the present to 0.6 Gyr ago show larger variations
among the different fields. There is evidence for a recent increase in
the SFR compared to the average past SFR in the H~IV, H~VI and H~VII
fields in which the star formation rate is about a factor of
$\approx3-4$ higher than the past star formation rate. On the other
hand, the recent SFR in the H~VIII and C25 fields is similar to or
somewhat less than the past average rate.  In fact the recent star
formation in the C25 field is partly enhanced by OB association No. 15
from the identifications of \citet{hod77} which happens to fall
partially on the WF4 chip in this field. Nevertheless, there are main
sequence stars in this field that lie outside the OB association.
These results confirm the conclusions of earlier ground-based
observations that the sites of recent star formation in NGC 6822 are
mostly confined to the bar while the older stars share a much wider
distribution throughout the galaxy  \citep[e.g.][]{hod77,hod91}.
Furthermore, the differences in the ratio of current (0.6 Gyr ago to
the present) SFRs to the older SFRs for the five fields
are consistent with a picture in which stars that formed as long ago
as 0.6 Gyr are not yet spatially very well mixed.

It is interesting to compare these results to the results derived from
previous ground-based data.  In particular, \citet{gal96b} analyzed
the SFH for the last 400 Myr and found evidence for an enhancement in
the SFR throughout the galaxy in the last 100-200 Myr.
The enhanced SFR was about a factor of two higher than the quiescent
SFR in the central area of the bar as well as in the outer regions
whereas the enhancement was found to be a factor of 4-6 in the regions
at the northern and southern ends of the bar.  I find an enhancement
in the H~IV field of a factor of 1.6 in the SFR about 200 Myr ago
compared to the SFR for 200-600 Myr ago. For the H~VI and H~VIII
fields I find a small enhancement in the most recent age bin while the
SFRs determined here in the H~VII and C25 fields are relatively
constant from 600 Myr ago to the present.  The most likely explanation
for these differences between the {\it HST} and ground-based data is
the small {\it HST} field-of-view. Apparently, the stars formed in the
SFR enhancement detected by \citet{gal96b} are still spatially clumped.

The recent star formation history for each field derived from the main
sequence luminosity function (see \S3.3) is plotted in Figure
\ref{recentsfh} as the solid red histogram while the dotted line
shows the recent SFRs from Table \ref{sfhtab2}.  The H~VI and
H~VII fields in the center of the bar show the highest SFR per area
while the recent SFR in the C25 field is the lowest. Except for the
outer C25 field, all of the recent SFHs show a small increase in the
SFR starting about 200 Myr ago with a further increase detected in the
H~IV, H~VI and H~VII fields in the last 100 Myr. These trends are
generally consistent with the the full SFH solutions. There is a
tendency for the SFRs derived from the main sequence luminosity
function to be smaller than the SFRs from the full fits to the CMD for
ages $>200$ Myr. This may be due in part to the selection of the
region of the CMD used to construct the  MS luminosity function. This
limit was chosen to include as many MS stars as possible while
minimizing the contamination by evolved stars.  Due to the increase in
photometric errors at fainter magnitudes (older ages) on the MS, some
true MS stars may lie to the red of the adopted MS boundary, thus
causing an underestimate of the SFR. The artificial star tests should
account for this effect to some degree.

\subsection{Extinction}

The best-fitting values of the extinction in each field are listed in
Table \ref{sfhtab1}. These values are mainly determined by the color
of the main sequence which is redder for those fields located within
the bar compared to the fields sampling more of the outer regions.
These values may be affected to some degree by the 
value of the final metallicity as discussed in \S4.3.
The extinction in the outer C25 field is $A_V=0.72$, or $E(B-V)=0.23$
while the extinction in the bar fields H~IV, H~VI and H~VII reaches
larger values of $A_V=1.04-1.09$, or $E(B-V)=0.34$, likely indicating
some amount of extinction internal to NGC 6822.  The extinction
towards the C25 field is consistent with the Galactic foreground
reddening of $E(B-V)=0.22-0.26$ derived from the Galactic 100 $\mu$m
dust emission \citep{sch98}. Similar results were also reached by
\citet{kay67} who found $E(B-V)=0.27\pm0.03$ based upon $UBV$
photometry of foreground stars in the direction of NGC 6822.  In
contrast to these results, reddenings based upon OB stars in NGC 6822
yield somewhat higher values of $E(B-V)=0.42\pm0.05$ \citep{vdb79} and
$E(B-V)=0.45$ \citep{wil92b}. The observations of \citet{mas95}
indicate an increase in the reddening from values of $E(B-V)=0.26$ in
the outer regions of NGC 6822 to values of $E(B-V)=0.45$ near OB stars
in the bar. The extinction values derived from the {\it HST}  CMDs are
consistent with these results although the $A_V$ values measured here
in the bar fields are not quite as high as the highest values found by
\citet{mas95}.  The discrepancy may be due to the fact that the
\citet{mas95} values are measured for a handful of massive individual
stars while each of the extinctions in Table \ref{sfhtab1} represents
an average over the entire WFPC2 field-of-view.

In Figure \ref{irasmap}, the locations of the five WFPC2 fields are
shown overlayed on the IRAS $60\mu$m map of NGC 6822 from
\citet{ric93}.  The corresponding value of $A_V$ derived from the CMD
fits is shown next to the location of each field.  While there is a
general correspondence between the locations of FIR peaks and the
luminous \ion{H}{2} regions, these sources only account for about
$50\%$ of the total FIR emission in NGC 6822 with the remainder coming
from a more diffuse component \citep{gal91,isr96}.  As \citet{gal91}
argued, the FIR is strong only in those locations where there is both
a high UV radiation field and a significant amount of dust. Thus, one
would expect there to be more FIR emission coming from those regions
with both higher extinction and a high density of young, massive
stars.  A comparison of the extinction value for each field with the
FIR flux at the same location is consistent with this expectation. The
C25 field falls within an area with little FIR emission, consistent
both with the low extinction measured for this field and the low
current SFR there. The FIR flux is much higher within the H~IV, H~VI
and H~VII fields where the CMDs indicate higher extinctions and higher
recent SFRs. Although the extinction and recent SFR derived from the
H~VIII field is only slightly larger than for the C25 field, there is
substantial FIR flux in this area. However, the FIR peak within the
H~VIII field is coincident with H~X,  perhaps indicating that most of
the FIR emission in this area is coming from this \ion{H}{2}
region. In fact, \citet{ode99} found that the CMD of the stars in H~X
is consistent with a reddening of $E(B-V)=0.35$ ($A_V=1.1$), a value
somewhat higher than the value of $A_V=0.82$ derived here for the
field near H~X.  This difference would be consistent with most of the
dust in this field being mostly in the vicinity of the \ion{H}{2}
region H~X.

\subsection{Comparison of Model and Observed Color-Magnitude Diagrams}

Using the model SFHs plotted in Figures \ref{sfh1} and \ref{sfh2},
model CMDs for each field were constructed and are plotted together in
Figure \ref{modelcmds} which can be directly compared to the observed
CMDs in Figure \ref{obs_cmds}.  Another method of displaying a CMD is
the Hess diagram, which is a surface that indicates the number of
stars within a series of color and magnitude bins in the CMD.  Hess
diagrams of each of the observed and model diagrams were constructed
for each field using the same CMD bins used in the SFH solution.  In
each panel of Figure \ref{hess}, the difference between the model and
observed Hess diagrams divided by the model diagram is shown.  Each
diagram is displayed from a ratio of $-1.5$ (black) to 1.5 (white).

Overall, the model CMDs reproduce the general morphology of the
observed data. There are important differences, however. Generally,
the main sequence in all of the fields is somewhat broader than
predicted simply based upon the results of the artificial star
tests. This may be due to my assumption of a single extinction in each
field when in reality there may be significant variation in the
extinction along the line-of-sight through NGC 6822. This
interpretation is supported by the tendency for there to be more
observed stars than predicted redward of the model main
sequence. Another possibility is that binaries may cause the main
sequence to be somewhat broader than predicted. A second difference
apparent between the models and observations is the distribution of
stars in the red clump.  The observed red clump is tilted upward in
the diagram. This results in an underabundance of model stars
brightward and blueward of the model clump while the models have too
many stars on the red and faint end of the clump. A similar
discrepancy is apparent in the SFH fits for the dwarf galaxy WLM
\citep{dol99}. One possible explanation could be differential
reddening within the field since the shape of the red clump is roughly
parallel to the reddening vector. In addition to these differences,
the asymptotic giant branch (AGB) population is evident in the models
CMDs extending brighter and redder of the RGB and is absent in the
observed CMDs.  However, the saturation limit of $I \approx 19$
renders much of the AGB unmeasured in the observations.

A fourth difference between the models and observations is apparent in
the area of the ``blue loops'' between the RGB and the main
sequence. These are evolved intermediate to massive stars undergoing
core He fusion. As is apparent from all the model CMDs, these stars
tend to lie along two sequences, one to the red and the other to the
blue. The magnitude of stars on both sequences increases with
decreasing mass with both sequences eventually merging into the
RC. The stars initially start on the red sequence and then loop over
to the blue sequence where they spend most of the remainder of their
He-burning lifetime. The amount of time stars spend on the blue loops
as well as the details of the evolution depend upon a number of
somewhat uncertain physical details, such as the nuclear reaction
rates and the extent of the convective core, among other details
\citep{chi92}. In addition, the color of the blue end of the loops
increases with metallicity with the loops disappearing near solar
metallicity \citep{ber94}. In the H~VI and H~VII fields there is a hint
of a sequence of red supergiants marking the red end of the blue loops
while the blue end of the loops, if present, is even weaker.  The
relative lack of identifiable sequences in the observed CMDs may be
due to some combination of effects including differential extinction,
binaries and uncertainties in the input physics to the evolutionary
models.  In addition, there are likely a number of foreground Galactic
stars contaminating this portion of the diagram. A similar situation
is apparent in the ground-based CMDs analyzed by \citet{gal96a}. In
their CMD, the red end of the loops is clearly visible while the blue
end of the loops appears to merge with the main
sequence. \citet{gal96a} interpreted this as evidence for differential
extinction within their data. 

\subsection{Discussion}

It is important to point out that the ages and SFHs derived in the
previous sections are based, among other things, on the assumption
that the distance to NGC 6822 is 500 kpc. 
This distance was determined from
observations of the Cepheids and assuming a distance modulus to the
LMC of $(m-M)_0=18.5$.  However, the distance modulus to the LMC
remains controversial. Van den Bergh (2000) found that an average of 18 of
the most recent distance measurements yields a value of $(m-M)_0=18.49$
while the individual determinations range all the way from 18.1 to
18.7. In addition, the implications of the {\it HIPPARCOS} mission on
the distance scale remain controversial.  (See discussions in
van den Bergh 2000 and Walker 1999). 

Changing the assumed distance to NGC 6822 would have an effect on the
intermediate and old SFH because shifting the inferred limits of the
RGB bins shown in Figure \ref{cmdregions} would change the color of
the red edge of the RGB.  Using the methods presented in \S3.2, this
might lead to a younger or older age for the oldest stars. However, as
can be seen in Figure \ref{h8timodel}, the shape of the RGB is also
affected by the age of the oldest stars. The RGB width and shape are
of course not affected by changes in the distance.  Based upon tests
of his SFH-fitting method on artificial data for a model galaxy,
\citet{dol97} found that errors in the distance modulus tended to
affect the SFRs determined for the older stars while leaving the
recent SFRs relatively unaffected.  For an error in the distance
modulus of $\pm0.06$ mag, \citet{dol97} determined that the SFRs in
the 5-10 and 10-15 Gyr age bins could systematically change by as much
as 40\%. The SFR averaged over the entire 5-15 Gyr age range though
was only off by at most 10\%. Therefore, the results presented here
for the age of the oldest stars and the SFHs would probably not be
affected to a large degree if the distance to the LMC were revised
somewhat.

From the SFHs presented in \S4.2, it is apparent that the SFH shapes
among the five fields are similar for older ages but diverge
within the last 0.5 Gyr, implying that stars formed with ages up
to 0.5 Gyr are not well mixed throughout the galaxy.
This mixing time scale of $\sim0.5$ Gyr derived from the SFH solutions can
be compared with other dynamical time-scales in NGC 6822. The velocity
dispersion of the \ion{H}{1} gas in NGC 6822 varies between about 7-10 km/s
across its major axis \citep{deb00}.  Assuming that stars are formed
with a similar velocity dispersion, the implied time scale is given by
\begin{equation}
\frac{D}{\sigma_{HI}} = \frac{2.9~{\rm kpc}}{7~{\rm km/s}} = 0.4~{\rm Gyr}
\end{equation}
where $D$ is the optical diameter of the galaxy. A second relevant
time scale is the rotation period as measured from the \ion{H}{1} gas. At a
radius of 1.4 kpc, the rotation speed is about 35 km/s
\citep{got77,deb00} which corresponds to  a rotation
period of 0.3 Gyr. Both of these time scales are similar to the mixing
times implied by the differences in the recent SFRs among the five HST
fields analyzed here.

The enhancement in the SFR 100-200 Myr ago detected in the
ground-based data of \citet{gal96b} was also detected in perhaps two
out of the five HST pointings analyzed in this paper. In addition to
the evidence from ground-based and {\it HST} CMDs, a period of
increased activity in NGC 6822 around this time is also supported by
the increase in the cluster formation rate 75-100 Myr ago detected by
\citet{hod80}. The low N/O ratio in NGC 6822 is consistent with a
recent episode of increased star formation as well \citep{ski89b}.
Oxygen is a primary nucleosynthesis product produced in massive stars
while nitrogen may be produced primarily as a secondary
nucleosynthesis product via the CNO cycle in intermediate mass stars
\citep[e.g., see discussion in ][]{smi75}. As pointed out by
\citet{ski89b}, for  example, the low $N/O$ ratio in NGC 6822 points
to a scenario in which a recent increase in the SFR
enriched the ISM with oxygen. If the increase happened recently
enough, there may not have been enough time for intermediate mass
stars to leave the main sequence and elevate the nitrogen abundance.

As NGC 6822 occupies a relatively isolated position in the Local Group
\citep[e.g.][]{mat98}, interactions with other known galaxies are not
the likely cause of this recent SFR enhancement. De Blok \& Walter
(2000) discussed the possibility that interaction between the main
body of NGC 6822 and the \ion{H}{1} cloud in the extreme NW of the
galaxy may be responsible for the increased star formation. According
to their data, the NW \ion{H}{1} complex has a total \ion{H}{1} mass
of $\sim 1.4 \times 10^7~M_{\odot}$. In addition, the \ion{H}{1}
velocity undergoes a sharp jump at the boundary between the cloud and
the rest of NGC 6822, perhaps indicating that it is actually a
separate system.  Furthermore, one-half the rotation period of the cloud
is $\sim 300$ Myr, a time scale comparable to the time-scale derived
from the recent SFH.

The recent SFHs derived from the {\it HST} CMDs can also be compared
with estimates of the current SFR from both the H$\alpha$ and 
far infra-red (FIR)
emission of NGC 6822. The total H$\alpha$ luminosity due to the
cataloged \ion{H}{2} regions in NGC 6822 is $3.0\times10^{39}$ ergs s$^{-1}$
\citep{hod89}
for a distance of 500 kpc and reddening of $E(B-V)=0.24$.
This implies a current SFR of $\geq 3\times10^{-9}$
M$_{\odot}$  yr$^{-1}$ pc$^{-2}$ where I have adopted the
SFR(H$\alpha$) calibration of \citet{ken98} and assumed a circular
area for NGC 6822 with radius 1.4 kpc. The SFR derived from the
H$\alpha$ luminosity is a lower limit because it excludes any possible
diffuse H$\alpha$ emission that may be ionized by Lyman continuum
photons that escape \ion{H}{2} regions and whose contribution should be
included when calculating the current SFR. On the other hand, the FIR
luminosity of NGC 6822 at wavelengths of $60-100$ $\mu$m is
$5\times10^7 L_{\odot}$ \citep{gal91}.  Under the assumption that the
FIR emission in galaxies is due to dust heated by stars with masses $M
\geq 2 M_{\odot}$ and that the stellar luminosity is converted into
$L_{FIR}$ with some efficiency $\eta$, the SFR averaged over the past
Gyr can be calculated from $L_{FIR}$ \citep{hun89}.  \citet{gal91}
argued that $\eta \leq 0.5$, implying that the recent SFR in NGC 6822
is $\geq 0.04$ M$_{\odot}$ yr$^{-1}$, or $\geq 7\times10^{-9}$
M$_{\odot}$ yr$^{-1}$ pc$^{-2}$, averaged over the same area as for
the H$\alpha$ SFR. Both the H$\alpha$ and $L_{FIR}$ SFR calibrations
assume a Salpeter IMF extending from 0.1 to 100 M$_{\odot}$.
These lower limits for the average recent SFRs
derived from the H$\alpha$ and FIR emission are comparable to the SFR
per area in the C25 and H~VIII fields while the other three {\it HST}
fields have somewhat higher SFRs per area over the past 200 Myr. 

There are a few other isolated dwarf irregular galaxies in the Local
Group besides NGC 6822 whose star formation histories have been
studied in similar detail. All of the evidence suggests that IC 1613
($M_B=-14.6$) has had an evolutionary history not radically different
from that of NGC 6822. RR
Lyrae stars have been detected in IC 1613, indicating that star
formation began in this galaxy at least 10 Gyr ago \citep{sah92}.
Based upon $VI$ CMDs from {\it HST}, \citet{col99} detected the presence
of a horizontal branch indicative of an old stellar
population. \citet{col99} also detected a number of asymptotic giant branch
stars and a strong RC implying that there is a strong intermediate age
(1-10 Gyr old) population in IC 1613 as well. While
the dwarf galaxy WLM ($M_B=-13.8$)
has a single globular cluster with an age of about
15 Gyr, it most likely began forming significant numbers of field
stars $\sim 12$ Gyr ago with about 50\% of its
stars formed 9-12 Gyr ago \citep{hod99,dol00a}. Although the
errors are large, star formation may have decreased 2.5-9 Gyr ago
before increasing again in the central regions within the past Gyr 
\citep{dol00a}. Based upon an analysis similar to that used by
\citet{gal96a} to study NGC 6822, \citet{apa97} found
that their ground-based CMDs of the dwarf galaxy Pegasus were best
fit with star formation beginning about 15 Gyr ago and then remaining
constant or perhaps gradually declining from then to the present.
Using CMDs from {\it HST}, \citet{gal98} argued that the star formation
rate in this galaxy was about a factor of 3-4 higher about a Gyr ago than
at present. Despite the differences in the their total luminosities, 
these three isolated dwarf irregulars are all similar to 
NGC 6822 in that they seem to have formed stars at more or less a
constant rate throughout their lifetimes with no evidence for very strong
variations greater than a factor of $\approx 3-4$.
All three galaxies are also similar to NGC 6822
in that they most likely began forming stars about 12-15 Gyr ago.
The one Local Group dwarf irregular galaxy whose star formation
history is an exception to the others is Leo A. Based upon the
distribution of stars in the RC and blue loops in the CMD,
\citet{tol98} argued that a major episode of star formation occurred in
Leo A 0.9-1.5 Gyr ago. Furthermore, these authors concluded that at
least 90\% of the stars in Leo A formed in the past 2 Gyr.

The SFHs for NGC 6822 derived here along with a
stellar population synthesis code can be used to infer the integrated
absolute magnitude of NGC 6822 as a function of time. 
Assuming a Salpeter IMF between 0.1 and 120 M$_{\odot}$, no internal
extinction and that the SFH derived from the H~VIII field is
representative, I have used the galaxy evolution code PEGASE
\citep{fio97} to determine the absolute $B$ and $V$ magnitudes of NGC
6822 as a function of look-back time. For an assumed SFH,
metal enrichment history, IMF and internal
extinction, the PEGASE code calculates the integrated spectrum for
both the stars and gas as a function of time, assuming a total mass of
stars formed of 1 M$_{\odot}$. This spectrum is then convolved with a
list of filter transmission curves to determine the magnitudes and
colors of the model galaxy in each of the filters.  I have scaled the
normalized magnitudes for the H~VIII field SFH by requiring that the
present $V$ absolute magnitude of the model galaxy match the observed
value of $M_V=-16.1$. 

The results are plotted in Figure \ref{mtot} in which the rest-frame
$B$ and $V$ absolute magnitudes for NGC 6822 are plotted as a function
of look-back time from the present. The overall trend is a gradual
increase in the luminosity of the galaxy with time.  The increase in
absolute magnitude about 4 Gyr ago corresponds to the drop in the SFR
around this time.  As discussed in \S4.2 and illustrated in Figure
\ref{h8tbins}, this drop in the SFR may not be very significant given
the similarity of the CMDs for stars between 1 and 7 Gyr old.  For
comparison the corresponding redshift is plotted on the upper axis
assuming a flat, cosmological constant dominated  cosmology with a
Hubble constant of $H_0=75$ km s$^{-1}$ Mpc$^{-1}$ and
$\Omega_{\Lambda}=0.7$. In the $I$-band selected CFRS redshift survey
\citep{lil95}, the faintest galaxies have $M_{AB}(B)=-16$, compared to
the current absolute magnitude of NGC 6822 of $M_{AB}(B)=-15.6$.
Galaxies with similar masses and evolutionary histories to NGC 6822
are therefore not likely  to be contributing to the evolving
population detected in the CFRS survey.  In the $b_j$ selected Autofib
redshift survey \citep{ell96}, the galaxies span the apparent
magnitude range $11.5 < b_j < 24.0$ and have redshifts in the range
$0<z<0.75$. Galaxies like NGC 6822 would be detectable in the Autofib
data out to redshifts of $z\approx0.2$. Assuming the $b_j$
K-corrections of \citet{kin85} for an Sdm galaxy and the same
cosmology as in Figure \ref{mtot} ($H_0=75$ km s$^{-1}$ Mpc$^{-1}$,
$\Omega=1$, $\Omega_{\Lambda}=0.7$), NGC 6822 would fade to
$b_j\approx27.4$ mag at $z=0.75$. Hence, galaxies similar to NGC 6822
are not likely contributing to the faint blue galaxy excess in either
the Autofib or CFRS redshift surveys.

The one caveat to this conclusion is that the age resolution of the
SFH solutions presented here decreases dramatically with age. If most
of the star formation at older ages occurred in one or more
short bursts, it may be possible that a galaxy such as NGC 6822 would
become momentarily bright enough to be detected in these redshift
surveys before fading again. As an example, I have recalculated the
evolution of the integrated absolute magnitude of NGC 6822 as a
function of age with the PEGASE code assuming that all of the stars
formed in the $5-7$ Gyr time bin were formed in a single star
formation burst of varying duration centered at an age of 6 Gyr. The
results are that NGC 6822 would temporarily brighten during the burst
to an absolute magnitude of $M_B=-19.5$, $-17.7$ and $-16.7$ for star
formation bursts of duration 10, 100 and 500 Myr, respectively.  In
the redshift range $0.50<z<0.75$, the CFRS detected galaxies down to
$M_{AB}(B)\simeq-19.5$. Since an age of 6 Gyr corresponds to
$z\approx0.6$, a galaxy similar to NGC 6822 would only be detectable
if all of the star formation in the $5-7$ Gyr time bin were squeezed
into a 10 Myr long star formation burst. At $z=0.6$,  the
corresponding apparent $b_j$ magnitudes would be 22.8, 24.6 and 25.6
for the 10, 100 and 500 Myr bursts, respectively.  Since the Autofib
survey included galaxies down to $b_j=24$, only a galaxy undergoing a
star burst of duration 10 Myr would be bright enough to make it into
that survey as well.

If there were such an intense burst of star formation in NGC 6822
$5-7$ Gyr ago, it would be difficult to detect with the CMDs presented
in this paper due to the relatively poor age resolution of the RC and
RGB structures in these diagrams where stars with these ages would be
found. The main sequence turn-off for 6 Gyr old stars occurs at
$M_V\approx3.5$ \citep{ber94}, which at the distance and extinction of
NGC 6822 corresponds to an apparent magnitude of $V\approx27.7$, a
level clearly much fainter than the current data.  Without deeper
photometry reaching much farther down the main sequence, it is not
possible to determine if such a star formation burst occurred in NGC
6822 or not. The recent enhancement by a factor of 2-4 in the SFR
100-200 Myr ago detected by \citet{gal96b} argues that the SFR in NGC
6822 likely has varied by similar amounts throughout its
history. Nevertheless, even if this is the case, variations of that
size would not be enough to render galaxies like  NGC 6822 detectable
in the current generation of redshift surveys.

\section{Summary}

This paper presents the first study of the star formation history
of the dwarf irregular galaxy NGC 6822 based upon {\it HST} CMDs
of five fields. The main conclusions of this study are summarized
below.

1) The best-fitting value of the extinction in the outer-most C25 field
is $A_V=0.72$, a value consistent with the expected amount of Galactic
foreground extinction toward NGC 6822. In the H~IV, H~VI and H~VII
fields in the bar, the extinction is $\approx0.3-0.4$ magnitudes
higher, presumably due to absorbing dust internal to NGC 6822.

2) The distribution of stars along the RGB implies that star formation
in NGC 6822 began 15 Gyr ago.  This conclusion rests upon two critical
assumptions: 1) that stars began forming from low-metallicity gas, as
indicated by the age and metallicity of the cluster H~VII, and 2) that
there is no spread in metallicity at a given age. If the initial
metallicity were higher or if the metal enrichment were rapid, then
the CMD could be fit by star formation beginning at younger ages.
Metallicities derived from spectra of individual RGB stars would be
useful in constraining the early chemical enrichment history of NGC
6822.

3) Using the methods developed by \citet{dol97}, the SFH in each field
was extracted. The overall shape of the SFH is quite similar among
the five fields with a higher overall SFR per area in the bar fields
compared to the C25 and H~VIII fields covering more out-lying regions.
In all five fields, except for a possible dip in the SFR from 3 to 5 Gyr
ago, the SFR has been fairly constant or perhaps somewhat increasing from 
15 Gyr ago to the present. For the H~VIII field, the color distribution
of stars along the RGB was used to constrain the SFH via the method
developed by \citet{gal96a}. The results are similar to the conclusions
drawn from the \citet{dol97} method.
A SFH in which star formation begins 12-15
Gyr ago and is either constant or undergoes a factor of four decrease
about 5 Gyr ago is best able to fit the observed RGB.

4) While the old and intermediate age SFHs of all five fields are similar,
the ratio of the recent SFR ($<0.6$ Gyr old) to the average past rate
varies among the five fields. The H~IV, H~VI and H~VII fields in the
bar show an increase by about a factor of 3-4 in the SFR in the past 0.6
Gyr compared to the average past rate. The remaining two fields that cover
more of the outer regions of the galaxy show a SFR that is roughly the
same or somewhat less than the average past rate. These results imply
that stars less than about 600 Myr old are not spatially well-mixed
throughout the galaxy. This is consistent with a scenario in which stars
form in OB associations and then slowly diffuse throughout the galaxy.
The time-scale derived from the analysis of the CMDs is similar to
the dynamical time-scales estimated from the rotation and velocity
dispersion of the \ion{H}{1} gas.

5) Model CMDs constructed from the best-fitting SFHs match the overall
morphology of the observed diagrams although there are differences
mainly in the area of the red clump and blue loops where the stellar
evolutionary models are more uncertain. The observed main sequence is
also somewhat broader in the observations compared to the models,
particularly in the bar fields. This may be due to the presence of
binary stars or differential extinction within each field which are
both not included in the models.

6) The SFH for the H~VIII field was used as input to the PEGASE
\citep{fio97} galaxy evolution code to infer the absolute $BV$
integrated magnitude of NGC 6822 as a function of look-back time and
redshift. In the rest-frame $B$ and $V$ bands NGC 6822 gradually
brightens with decreasing redshift except for a temporary decrease in
the luminosity during the lull in star formation 3-5 Gyr ago. For a
flat cosmology with  $H_0=75$ km s$^{-1}$ Mpc$^{-1}$ and
$\Omega_{\Lambda}=0.7$, galaxies with evolutionary histories similar
to NGC 6822 would be undetectable in the CFRS redshift survey
\citep{lil95} and would be included in the Autofib survey
\citep{ell96} only out to a redshift of $z\approx0.2$. However, if
most of the stars older than a Gyr were formed in a single short burst
of star formation, such a galaxy might then have been bright enough to
become detectable in the current generation of redshift surveys. In
particular, if all of the star formation from 5 to 7 Gyr ago were
compressed into a burst lasting only 10 Myr, a galaxy similar to NGC
6822 would temporarily rise above the detection limits of both the
CFRS and Autofib surveys.

\acknowledgements{I am grateful to Andrew Dolphin for providing me
with his HSTPHOT/MULTIPHOT photometry package as well as many of the
computer programs used in deriving the star formation histories
presented in this paper. I also wish to thank Julianne Dalcanton and
Eugene Magnier for helpful comments. I am particularly grateful to my
thesis advisor Paul Hodge for all his excellent help and guidance
throughout this project. I also wish to thank the anonymous referee
for carefully reading the manuscript and for making many insightful
comments which have significantly improved this paper.  Partial
support for this work was provided by NASA through grant numbers
GO-08314 and AR-08362 from the Space Telescope Science Institute,
which is operated by the Association of Universities for Research in
Astronomy, Incorporated, under NASA contract NAS5-26555.}

\newpage

\figcaption{Digitized Sky Survey image of NGC 6822. The entire
field-of-view is $15\arcmin \times 15\arcmin$ with North up and East
to the left. The positions of the five WFPC2 fields analyzed here
are shown. Each field is labeled by the cluster or \ion{H}{2} region
that is centered on the PC chip. The cluster C25 is from the list
of clusters in \citet{hod77} while the remaining objects labeled with
an H were originally identified by \citet{hub25}.\label{dss}} 

\figcaption{Observed $V$, $V-I$ CMDs for all stars detected in the
WF chips of each field. The contour levels correspond to stellar densities
of 20, 40, 60, 80, 100, 150, 200, 300, 400, and 500 stars decimag$^{-2}$.
For those areas of each CMD with fewer than 20 stars decimag$^{-2}$, each 
individual star is plotted. See Figure \ref{dss} for the positions of these
fields in NGC 6822. \label{obs_cmds}}

\figcaption{Fraction of the artificial stars recovered in
both $V$ and $I$ as a function of the input $V$ magnitude for all five
fields.  Each panel is labeled with the corresponding field name. The
solid, dotted and dashed lines are the completeness curves for
artificial stars with input $(V-I)$ colors of $-0.5$ to 0.5, 0.5 to
1.5, and 1.5 to 2.5, respectively.\label{fakestars}}

\figcaption{The fraction of stars from the H~VII photometry that were
also detected in the H~VI field as a function of $V$ magnitude (solid
line).  Only stars in the H~VII field that lie within the
field-of-view of the H~VI field were included in the calculation. The
dotted line shows the probability that a star should be detected on
both fields based upon the results of the artificial star tests. This
prediction is simply the product of the completeness fractions for the
H~VI and H~VII fields for artificial stars with input colors
$0.5<V-I<1.5$. The agreement between the observed and predicted
fractions argues that the artificial star tests provide an accurate
description of the completeness.\label{h6vsh7_fraction}}

\figcaption{(a) The expected error in the $V$ photometry $\sigma_V$ derived
from the artificial star tests for the H~VI field (pluses) and H~VII
field (diamonds). (b) The difference in $V$ magnitude $\Delta V$ divided
by the expected error as a function of magnitude for stars detected
in both the H~VI and H~VII fields. The differences are in the sense of
the H~VI minus the H~VII photometry. Panels (c) and (d) are the
same as (a) and (b) except the $I$-band results are plotted.
\label{h6vsh7}}

\figcaption{The three chemical enrichment laws used as the input for
each SFH solution using the \citet{dol97} method for an initial
metallicity of $[{\rm Fe/H}]=-2.3$ and a final metallicity of $[{\rm
Fe/H}]=-0.7$. The solid, dotted and dashed curves in this figure are
referred to in the text as chemical enrichment laws (CELs) 1, 2 and 3,
respectively.
\label{fz}}

\figcaption{Model CMDs, separated by age, generated from the H~VIII
field artificial star test results. These CMDs assume a constant SFR
of $5 \times 10^{-4}$ $M_{\odot}$ yr$^{-1}$, an initial metallicity of
$[{\rm Fe/H}]=-2.3$ and a final metallicity of $[{\rm Fe/H}]=-0.7$ and
chemical enrichment law 2, as plotted in Figure \ref{fz}.
\label{h8tbins}}

\figcaption{Three different SFH solutions for the H~VIII field
assuming three different sets of time bins. In panel (a), the results
for Test 1 are shown for all 12 time bins used in the solution. In
panel (b), the same solution is shown, rebinned to match the time bins
used in the results presented in \S4 of the text.  The results for Test
2 (32 time bins with spacing $\Delta {\rm log}(t)=0.1$) are shown
panels (c) and (d) while panels (e) and (f) show the results for Test
3 (16 time bins with spacing $\Delta {\rm log}(t)=0.2$).
\label{timeres}}

\figcaption{The five possible input star formation history shapes
tried. The youngest age considered has been set to $T_f=10$ Myr while
$T_i$ is the age of the oldest stars. For all shapes, except the
constant SFR case (Shape 3), the higher SFR is four times the  lower
SFR. Each of the time bins is $1/3$ of $T_i$ in length and all three
curves have been normalized to have the same integrated mass of stars
formed. These shapes are identical to those assumed by
\citet{gal96a}.\label{sfhshapes}}

\figcaption{The H~VIII CMD assuming a
distance modulus of 23.49 and extinction of $A_V=0.82$. The regions
marked with the numbers 1 to 4 indicate the regions of the red giant
branch used to compare the models and observations while the region
labeled RC is defined to be the red clump region.
All stars that lie blueward of the dashed line are assumed to be
main sequence stars when computing the main sequence luminosity function.
\label{cmdregions}}

\figcaption{Model RGB color distributions (solid lines)
assuming $Z_i=0.0001$ ($[{\rm Fe/H}]=-2.3$) for the stars within the
RGB region 1 shown in Figure \ref{cmdregions}. The number in the upper
left-hand corner of each panel indicates the SFH shape used in the
model. Each of the columns shows the model RGB color distributions
for four values of $T_i$ (from left to right): 15, 12, 9 and 6 Gyr.
In each panel the observed color distribution is plotted as the
dotted line.\label{rgbcdist}}

\figcaption{The fit parameter $\bar{\chi}^2$ as a function of the
time when star formation begins ($T_i$) for different assumptions
regarding the metal enrichment history. Each panel shows the
variation of the fit parameter with $T_i$ for a particular choice
of initial metallicity $[{\rm Fe/H}]_i$ and the final metallicity
$[{\rm Fe/H}]_f$ for the three different metal enrichment laws
shown in Figure \ref{fz}. The solid, dotted and dashed lines
correspond to the chemical enrichment laws 1, 2 and 3, respectively.
An extinction of $A_V=0.8$ mag was assumed for all of the solutions
plotted here.\label{h8tiz}}

\figcaption{Best-fit star formation and chemical enrichment histories
for the H~VIII field under three different assumptions of the age
of the oldest stars $T_i$. The results are shown for $T_i=15$, 12
and 9 Gyr in the top, middle and bottom panels, respectively.
\label{h8tisfh}}

\figcaption{Model CMDs for the H~VIII field under three different
assumption for the age of the oldest stars $T_i$: (a) 15 Gyr, (b) 12
Gyr and (c) 9 Gyr. The observed CMD is plotted in (d). In each panel
the contours correspond to densities of 20, 40, 60, 80, 100, 150, 200,
300 and 400 stars decimag$^{-2}$.  The SFHs and chemical enrichment
histories corresponding to these three model CMDs are plotted in
Figure \ref{h8tisfh}.\label{h8timodel}}

\figcaption{Indicators relative to the color distribution of stars in
the brightest RGB region in the H VIII field CMD as a function of the
star formation history shape assuming a low initial metallicity of
$[{\rm Fe/H}]=-2.3$ ($Z=0.0001$)  and a final metallicity of $[{\rm
Fe/H}]=-0.7$ ($Z=0.004$).  (See Figure \ref{cmdregions} for the
definition of the RGB regions). Note that the SFH shapes are ordered
such that they progress from a dominant young population in Shape 1 to
a dominant older population in Shape 5.  The median and C95 $V-I$
colors are plotted in panels (a) and (b). The ratio of the number of
RGB stars to red clump stars is plotted in (c) while (d) shows the
value of $1/\chi^2$ for each model. In each panel, the results are
shown separately for $T_i$ of 15 Gyr (dotted line), 12 Gyr (dashed
line), 9 Gyr (dot-dashed line) and 6 Gyr (long-dashed line). The solid
lines denote the observed value and its probable error due to an
uncertainty of $\pm 0.1$ mag in the extinction $A_V$.\label{cmdvar1}}

\figcaption{Same as Figure \ref{cmdvar1} except for an
initial metallicity of $[{\rm Fe/H}]=-1.0$.\label{cmdvar2}}

\figcaption{The star formation rate (left panels)  plotted in units of
$10^{-9}$ $M_{\odot}$ yr$^{-1}$ pc$^{-2}$ and chemical enrichment
history (right panels) for the H~VIII field under three different
assumptions about the extinction $A_V$. In the top, middle and bottom
panels, the extinction was held fixed at values of $A_V=0.6$, 0.9 and
1.2, respectively, while allowing the  initial metallicity, final
metallicity and enrichment law to vary.
\label{h8avsfh}}

\figcaption{The value of the fit parameter as a function of (a) the
extinction $A_V$, (b) the initial metallicity $[{\rm Fe/H}]_i$, (c)
the final metallicity $[{\rm Fe/H}]_f$ and (d) the chemical enrichment
law, as defined in Figure \ref{fz}. All SFH solutions plotted here are
fits to the H~VIII field CMD.\label{h8fit}}

\figcaption{The SFR as a function of age derived from the H~VIII field
CMD under three different assumptions about the chemical enrichment
law. The SFH for chemical enrichment laws 1, 2 and 3 are shown in the
top, middle and bottom panels, respectively. In each case, the initial
and final metallicities are set to $[{\rm Fe/H}]=-2.3$ and $-0.9$,
respectively.\label{h8amr1}}

\figcaption{Same as Figure \ref{h8amr1}, except assuming an initial
metallicity of $[{\rm Fe/H}]=-1.4$.\label{h8amr2}}

\figcaption{Star formation rate (left panels) and chemical enrichment
history (right panels) for the H~IV, H~VI and H~VII fields. The SFR
is plotted in units of $10^{-9}$ $M_{\odot}$ yr$^{-1}$ pc$^{-2}$,
averaged over the three WF chips. The metallicities are plotted
for the initial and final values in each time bin.\label{sfh1}}

\figcaption{Star formation rate (left panels) and chemical enrichment
history (right panels) for the H~VIII and C25 fields. The SFR
is plotted in units of $10^{-9}$ $M_{\odot}$ yr$^{-1}$ pc$^{-2}$,
averaged over the three WF chips.\label{sfh2}}

\figcaption{The SFH of NGC 6822
determined from the H~VIII field. The points with error bars are
the results determined using the \citet{dol97} method. The solid
and dotted lines correspond to the two best-fitting SFHs
determined from the \citet{gal96a} method. The solid line
corresponds to SFH shape 4 while the dotted line is SFH shape 3. 
Both solutions have $T_i=15$ Gyr and $Z_i=0.0001$.\label{h8sfh}}

\figcaption{Recent SFHs for all five fields derived from the main
sequence luminosity function (red solid histogram with error bars).
For comparison, the recent star formation rates derived from the
\citet{dol97} method are shown as the dotted line with error
bars.\label{recentsfh}}

\figcaption{The locations of the five WFPC2 fields overlayed on the
$60\mu$m emission measured by the IRAS satellite \citep{ric93}.  The
coordinates are epoch 1950 and  the contours correspond to surface
brightnesses of 1, 2, 3, 4, 5, 10, 15, 20, 25, 30 and 40 MJy
sr$^{-1}$. The value of the extinction $A_V$ determined from the CMD
is shown next to each field. The WFPC2 fields with higher extinction
tend to lie in areas of higher IRAS emission.\label{irasmap}}

\figcaption{Model CMDs reconstructed from the best-fitting SFHs
plotted in Figures \ref{sfh1} and \ref{sfh2}.
In all panels, the contours correspond to
densities of 20,40,60,80,100,150,200,300,400 and 500 stars
decimag$^{-2}$.  (See Figure \ref{obs_cmds} for the observed
CMDs).\label{modelcmds}}

\figcaption{Grayscale Hess images showing the difference
between the observed and model Hess diagrams divided by the model
diagram for each of the five fields. In each panel white corresponds
to more model stars than observed stars. The ratio has been plotted from
a value of $-1.5$ (white) to $+1.5$ (black).\label{hess}}

\figcaption{Integrated absolute rest-frame
$B$ (solid line) and $V$ (dotted line) 
magnitudes for NGC 6822 as a function of look-back time. The curves were
calculated from the SFH for the H~VIII field using the PEGASE galaxy
evolution code \citep{fio97}. The upper x-axis shows the redshift
corresponding to each age on the bottom axis assuming a flat cosmology
with a Hubble constant $H_0=75$ km/s/Mpc and $\Omega_{\Lambda}=0.7$.
\label{mtot}}

\newpage

\begin{deluxetable}{lcccc}
\tablecaption{Summary of observations\label{obs_log}}
\tablewidth{0pt}
\tablehead{
\colhead{Field} & \colhead{File} & \colhead{Filter} &
\colhead{Exp. time (sec)} & \colhead{Date of obs.}
}
\startdata
H~IV & u37h0201r.c0h & F555W & 2600 & 1999 July 21 \\
    & u37h0202r.c0h & F555W & 1300 & 1999 July 21 \\
    & u37h0203r.c0h & F814W & 2700 & 1999 July 21 \\
    & u37h0204r.c0h & F814W & 1200 & 1999 July 21 \\
H~VI & u37h0301r.c0h & F555W & 2487 & 1999 March 24 \\ 
    & u37h0302r.c0h & F555W & 1300 & 1999 March 24 \\ 
    & u37h0303r.c0h & F814W & 2700 & 1999 March 24 \\ 
    & u37h0304r.c0h & F814W & 1200 & 1999 March 24 \\
H~VII & u37h0401r.c0h & F555W & 2600 & 1999 March 26 \\
     & u37h0402r.c0h & F555W & 1300 & 1999 March 26 \\
     & u37h0403r.c0h & F814W & 2700 & 1999 March 27 \\
     & u37h0404r.c0h & F814W & 1200 & 1999 March 27 \\
H~VIII & u37h0501r.c0h & F555W & 2600 & 1999 March 29 \\
      & u37h0502r.c0h & F555W & 1300 & 1999 March 29 \\
      & u37h0503r.c0h & F814W & 2700 & 1999 March 29 \\
      & u37h0504r.c0h & F814W & 1200 & 1999 March 29 \\
C25 & u5ch0403r.c0h & F555W & 800 & 1999 September 24 \\
    & u5ch0404r.c0h & F555W & 400 & 1999 September 24 \\
    & u5ch0405r.c0h & F814W & 600 & 1999 September 24 \\
    & u5ch0406r.c0h & F814W & 600 & 1999 September 24 \\
\enddata
\end{deluxetable}

\begin{deluxetable}{llrrrr}
\tablecaption{Comparison of H VI and H VII photometry\label{h6vsh7tab}}
\tablewidth{0pt}
\tablehead{
\multicolumn{2}{c}{Chip} & \colhead{} & \colhead{} \\
\cline{1-2} \\
\colhead{H VI} & \colhead{H VII} & \colhead{$\Delta V$} & 
\colhead{$\sigma_V$} & \colhead{$\Delta I$} & \colhead{$\sigma_I$}
}
\startdata
  WF2 & PC  & $-$0.046 & 0.127 & $-$0.029 & 0.092\\
  WF2 & WF2 &    0.016 & 0.085 &    0.000 & 0.061\\
  WF3 & PC  & $-$0.050 & 0.190 & $-$0.076\tablenotemark{a} & 0.199\\
  WF3 & WF2 &    0.011 & 0.068 & $-$0.043 & 0.063\\
  WF3 & WF3 &    0.003 & 0.081 & $-$0.034 & 0.073\\
  WF3 & WF4 &    0.000 & 0.093 & $-$0.009 & 0.070\\
  WF4 & WF4 & $-$0.006 & 0.079 &    0.010 & 0.067\\
\enddata
\tablenotetext{a}{Average is affected by a few discrepant bright stars}
\end{deluxetable}

\begin{deluxetable}{llll}
\tablecaption{Parameters used to solve for the SFH\label{sfhpar}}
\tablewidth{0pt}
\tablehead{
\colhead{Parameter} & \colhead{Minimum} & \colhead{Maximum} & \colhead{Step}
}
\startdata
$A_V$ & 0.6 & 1.3 & 0.1\\
$[Fe/H]_i$ & $-2.3$ & $-1.4$ & 0.3\\
$[Fe/H]_f$ & $-1.1$ & $-0.7$ & 0.2\\
\cutinhead{Fixed Parameters}
$(m-M)_0$ & 23.49 & & \\
$\alpha$ & 1.35 & & \\
binary fraction & 0.0 & &  \\
$\sigma$([Fe/H]) & 0.0 & & \\
\enddata
\end{deluxetable}

\begin{deluxetable}{ccccc}
\tablecaption{SFH Solutions for the H~VIII Field for $T_i=15,11,8$ Gyr
\label{h8tab1}}
\tablewidth{0pt}
\tablehead{
\colhead {} & \multicolumn{2}{c}{Fit parameter} & \colhead{} & \colhead{} \\
\cline{2-3} \\
\colhead{$T_i$ (Gyr)} & \colhead{Min} & \colhead{Max} & {Num. of solutions} &
\colhead{$A_V$}
}
\startdata
15 & 3.19 & 3.50 & 30 & $0.817\pm0.072$ \\
12 & 3.25 & 3.65 & 32 & $0.824\pm0.076$ \\
9  & 3.35 & 3.85 & 32 & $0.839\pm0.076$ \\
\enddata
\end{deluxetable}

\begin{deluxetable}{cccc}
\tablecaption{SFRs for the H~VIII Field for $T_i=15,12,9$ Gyr\label{h8tab2}}
\tablewidth{0pt}
\tablehead{
\colhead{} & 
\multicolumn{3}{c}{SFR ($10^{-9}$ $M_{\odot}$ yr$^{-1}$ pc$^{-2}$)} \\
\cline{2-4} \\
\colhead{Age (Gyr)} & \colhead{$T_i=15$ Gyr} & \colhead{$T_i=12$ Gyr} &
\colhead{$T_i=9$ Gyr}
}
\startdata
$0.001-0.2$   & $6.1\pm0.7$   & $6.4\pm1.0$ & $6.4\pm1.0$ \\
$0.2-0.6$     & $4.7\pm3.6$   & $4.7\pm0.4$ & $4.8\pm0.3$ \\
$0.6-3$       & $7.3\pm0.9$   & $7.0\pm1.1$ & $6.8\pm1.6$ \\
$3-5$         & $0.05\pm0.16$ & $1.4\pm2.7$ & $6.0\pm3.6$ \\
$5-7$         & $6.2\pm3.5$   & $5.7\pm4.4$ & $0.7\pm2.1$ \\
$7-T_i$       & $2.6\pm1.7$   & $3.4\pm2.5$ & $6.4\pm4.9$ \\
\enddata
\end{deluxetable}

\begin{deluxetable}{ccccc}
\tablecaption{SFH solutions for the Five {\it HST} Fields\label{sfhtab1}}
\tablewidth{0pt}
\tablehead{
\colhead {} & \multicolumn{2}{c}{Fit parameter} & \colhead{} & \colhead{} \\
\cline{2-3} \\
\colhead{Field} & \colhead{Min.} & \colhead{Max.} &
\colhead{Num. of solutions} & \colhead{$A_V$}
}
\startdata
H IV   & 3.54 & 3.80 & 31 & $1.041\pm0.070$ \\
H VI   & 5.00 & 5.40 & 33 & $1.090\pm0.071$ \\
H VII  & 4.71 & 5.16 & 32 & $1.019\pm0.069$ \\
H VIII & 3.19 & 3.50 & 30 & $0.817\pm0.072$ \\
C25    & 2.28 & 2.80 & 28 & $0.720\pm0.073$ \\
\enddata
\end{deluxetable}

\begin{deluxetable}{cccccc}
\tablecaption{SFRs for the Five {\it HST} Fields\label{sfhtab2}}
\tablewidth{0pt}
\tablehead{
\colhead{} &
\multicolumn{3}{c}{SFR ($10^{-9}$ $M_{\odot}$ yr$^{-1}$ pc$^{-2}$)} \\
\cline{2-6} \\
\colhead{Age (Gyr)} & \colhead{H IV} & \colhead{H VI} & \colhead{H VII} &
\colhead{H VIII} & \colhead{C25}
}
\startdata
$0.001-0.2$ & $14.5\pm 1.7$ & $25.4\pm 3.4$ & $18.0\pm 2.6$ & $ 6.1\pm 0.7$
& $ 3.8\pm 0.3$\\
  $0.2-0.6$ & $ 8.8\pm 1.0$ & $21.4\pm 2.7$ & $18.3\pm 1.7$ & $ 4.7\pm 0.4$
& $ 3.4\pm 0.5$\\
    $0.6-3$ & $ 8.1\pm 1.0$ & $13.4\pm 2.2$ & $11.8\pm 1.7$ & $ 7.3\pm 0.9$
& $ 8.6\pm 1.2$\\
      $3-5$ & $ 0.0\pm 0.0$ & $ 0.1\pm 0.5$ & $ 0.0\pm 0.3$ & $ 0.1\pm 0.2$
& $ 0.2\pm 0.2$\\
      $5-7$ & $ 7.5\pm 3.4$ & $12.4\pm 5.7$ & $10.9\pm 4.6$ & $ 6.2\pm 3.5$
& $10.3\pm 4.8$\\
      $7-15$ &$ 3.7\pm 2.1$ & $ 4.1\pm 3.0$ & $ 7.6\pm 2.8$ & $ 2.6\pm 1.7$
& $ 1.4\pm 1.3$\\
\enddata
\end{deluxetable}

\end{document}